\documentclass[10pt,pra,aps,twocolumn,showpacs,amsmath,amssymb,superscriptaddress]{revtex4-1}
\usepackage{times}
\usepackage{textcomp}
\usepackage{graphicx}
\usepackage{bm}
\allowdisplaybreaks 
\usepackage{braket}
\usepackage[breaklinks]{hyperref}
\hypersetup{colorlinks=true, linkcolor=blue, citecolor=blue, filecolor=blue, urlcolor=blue}

\newcommand{\be}{\begin{equation}}
\newcommand{\ee}{\end{equation}}

\newcommand{\eea}{\end{eqnarray}}

\begin{document}

%%%%%%%%%%%%%%%%%%%%%%%%%%%%%%%%%%%%%%%%%%%%%%%%%%%%%%%%%%%%%%%%%%%%%%%%%%%%%

\title{Obstacles to quantum annealing in a planar embedding of XORSAT}

\author{Pranay Patil}

\affiliation{Department of Physics, Boston University, 590
  Commonwealth Avenue, Boston, Massachusetts 02215, USA\\}
  
\author{Stefanos Kourtis}

\affiliation{Department of Physics, Boston University, 590
  Commonwealth Avenue, Boston, Massachusetts 02215, USA\\}

\author{Claudio Chamon}

\affiliation{Department of Physics, Boston University, 590
  Commonwealth Avenue, Boston, Massachusetts 02215, USA\\}

\author{Eduardo R. Mucciolo}

\affiliation{Department of Physics, University of Central Florida,
  Orlando, Florida 32816, USA\\}

\author{Andrei E. Ruckenstein}

\affiliation{Department of Physics, Boston University, 590
  Commonwealth Avenue, Boston, Massachusetts 02215, USA\\}

%%%%%%%%%%%%%%%%%%%%%%%%%%%%%%%%%%%%%%%%%%%%%%%%%%%%%%%%%%%%%%%%%%%%%%%%%%%%%
\begin{abstract}
We introduce a planar embedding of the $k$-regular $k$-XORSAT
problem, in which solutions are encoded in the ground state of a
classical statistical mechanics model of
reversible logic gates arranged on a square grid and acting on bits
that represent the Boolean
variables of the problem. The special feature of this embedding is that
the resulting model lacks a finite-temperature phase transition,
thus bypassing the first-order thermodynamic
transition known to occur in the random graph representation of XORSAT.
In spite of this attractive feature, the thermal
relaxation into the ground state displays remarkably slow glassy
behavior.
The question addressed in this paper is whether this planar embedding
can afford an efficient path to solution of $k$-regular $k$-XORSAT
via quantum adiabatic annealing. We first show
that our model bypasses an avoided level crossing and consequent
exponentially small gap in the limit of small transverse fields. We
then present quantum Monte Carlo results for our embedding of the
$k$-regular $k$-XORSAT that strongly support a picture in which
second-order and first-order transitions develop at a finite
transverse field for $k=2$ and $k=3$, respectively. This translates
into power-law and exponential dependences in the scaling of energy
gaps with system size, corresponding to times-to-solution which are,
respectively, polynomial and exponential in the number of variables.
We conclude that neither classical nor quantum annealing can
efficiently solve our reformulation of XORSAT, even though the
original problem can be solved in polynomial time by Gaussian
elimination. \end{abstract}

\maketitle

%%%%%%%%%%%%%%%%%%%%%%%%%%%%%%%%%%%%%%%%%%%%%%%%%%%%%%%%%%%%%%%%%%%%%%%%%%%%%
\section{Introduction}
%Talk about general quantum annealing (find refs), 

Boolean satisfiability problems form an important subset of computer science
problems which have been studied extensively with regard to their general
solvability. These problems usually comprise of finding a set of bits, each of
which can take values of either 0 or 1, which satisfy the set of constraints
demanded by the particular problem. One of the simplest examples of such a
constraint satisfaction problem is XOR-satisfiability,
where each constraint (or clause)
reads in a set of bits and requires their sum modulo 2 to be 0 or 1. A bit
assignment (arrangement of 0's and 1's) which satisfies all the clauses in 
an XORSAT problem is said to be a solution.

Connections between such computer science (CS) problems
and statistical physics models of bits
(spins) is an active area of research. The generic approach encodes
the solution of the computational problem in the ground state of a
statistical mechanics model. This line of research has led to novel
insights and algorithms, such as simulated annealing and belief
propagation~\cite{Kirkpatrick1983,Zechina-Parisi-Mezard,Montanari-and-Mezard-book},
which have become standard tools in many applied science areas. A
shortcoming of most implementations of this approach is that the
underlying physical systems tend to display glass transitions that
prevent one from reaching the ground state, thus blocking the
path-to-solution for the computational problem. While one may be
tempted to connect the complexity of the computational problem to the
glassiness of the model, this naive expectation is misdirected. A
glass transition, and thus a barrier to solution, also occurs in some
of the current statistical mechanics realizations of XORSAT, an ``easy''
computational problem in class P~\cite{Ricci-Tersenghi2010}.

In a set of recent publications~\cite{Chamon2016,Lei} we introduced a
planar embedding of universal classical computation which displays no
finite temperature bulk phase transitions, thus eliminating an obvious
obstruction to reaching solution through thermal annealing. In spite
of this conceptual simplification, within our mapping the difficulty
is transferred to the dynamical behavior as one cools to low
temperatures. We find that, even for some simple computational
problems (such as multiplication) and 3-regular 3-XORSAT, a
problem which is solvable in polynomial time by Gaussian elimination
\cite{GE}, the relaxation times into the ground state and thus the
times-to-solution are extremely long. The source of glassiness in
our reformulation of
these easy computational problems can be traced back to two
qualitatively different mechanisms. In the case of multiplication, a
single thermally excited defect created in the annealing process
translates into a macroscopic number of computational errors
~\cite{Lei}. On the other hand, for 3-regular 3-XORSAT
our statistical mechanics representation involves partial knowledge of
input/output boundaries (see Appendix). For this ``mixed-boundary
conditions'' case we
have shown~\cite{Chamon2016} that thermal annealing is an
ineffective way of reaching solution (i.e., of determining the full
state of the boundaries).

In this paper, we explore the alternative of employing quantum adiabatic
annealing (QAA)~\cite{TNat,EFold,EF,QATIM,QAscaling,EFscience} to solve
our planar model of XORSAT efficiently. Specifically, we are motivated
to ask whether our reformulation, which removes the first-order thermal
transition, also removes the first-order quantum phase transition that
was encountered in the original quantum annealing formulation
of XORSAT on random regular graphs~\cite{EFold}.
We present quantum Monte Carlo results for
our embedding of the $k$-regular $k$-XORSAT. In particular, we study
the nature of the phase transition along the quantum axis as a
function of transverse field for $k=2$ and $k=3$. The first
obstruction one could encounter in approaching the classical ground
state along the quantum axis is an avoided level crossing, resulting in
an exponentially small gap in the limit of small transverse
fields and signaling an exponentially long time to
solution~\cite{Alt}. We show that our embedding does not suffer from
this avoided crossing shortcoming.

The $k=2$ case is simple and can be solved via both thermal and
quantum annealing. In particular, the $k=2$ model displays a
second-order phase transition along the quantum axis, corresponding to
a time-to-solution that scales polynomially with the system size. In
contrast, our results show that the $k=3$ model undergoes a
first-order transition at a finite transverse field. This translates
into an exponential scaling of energy gaps and thus to a
time-to-solution that scales exponentially with the number of
variables in the XORSAT instances.
Thus, even though our embedding bypasses two of the
obvious obstructions to annealing into the ground state --- the
classical phase transition at finite temperature along the classical
axis and the perturbation theory collapse for small transverse field
along the quantum axis --- our results for $k=3$ reinforce the
conclusion that both classical and quantum annealing can be ineffective
pathways to solution even for easy (i.e., complexity P) computational
problems. The same conclusion was reached by Farhi {\it et
  al}.~\cite{EFold}, whose approach also bypassed the ``perturbation
theory collapse'' of Ref.~\onlinecite{Alt} but not the thermodynamic
transition along the classical axis. In summary, our embedding of
classical computational problems into statistical mechanics models
defines a class of systems with short-range interactions and no
classical bulk thermal phase transitions that display slow glassy
dynamics in both thermal and quantum annealing.

The outline of the paper is as follows. In Sec.~\ref{sec:qaa} we
provide an introduction to quantum annealing in the context of the
general class of problems we are interested in studying, along with a
comparison to simulated annealing. This is followed by
Sec.~\ref{sec:3xor} where we describe the particular problem we study
and its lattice embedding that we investigate for efficiency of
annealing. In Sec.~\ref{sec:anres} we provide an interpretation for
the efficacy of simulated annealing in reaching solutions in two
classes of regular XORSAT problems, as well as weak perturbation
theory argument concerning quantum annealing at small transverse
fields. Sec.~\ref{sec:QMCres} presents numerical results
from quantum Monte Carlo for a simple illustrative example
of a single CNOT gate and builds on these results to address the more
complex case of $k=2,3$ regular XORSAT through similar simulations. We
present our conclusions in Sec.~\ref{Conc}.

%%%%%%%%%%%%%%%%%%%%%%%%%%%%%%%%%%%%%%%%%%%%%%%%%%%%%%%%%%%%%%%%%%%%%%%
\section{Quantum adiabatic annealing algorithm: outline and general considerations}
\label{sec:qaa}

The quantum adiabatic annealing (QAA) algorithm was introduced as a
method to solve computational problems in
Refs.~\cite{EFold,Farhi2001}. It exploits long-range quantum coherence
in a time dependent quantum system whose defining Hamiltonian interpolates
adiabatically between two limits. The goal is to use QAA to reach the
ground state of a complicated classical Hamiltonian $H$, which encodes
the solution of the computational problem at hand, by adiabatically deforming the quantum
ground state of a ``simpler'' initial Hamiltonian, which is easy to
prepare.

Concretely, a computational problem on $N$ Boolean variables is mapped
to a Hamiltonian $H$ that describes interactions between $N$ classical
Ising spin degrees of freedom $\sigma_i^z =
\ket{\uparrow},\ket{\downarrow}$ for $i=1,\dots,N$. The spin-up state
$\sigma_i^z =\ket{\uparrow}$ can be chosen to represent bit state
$x_i=1$ and the spin-down state $\sigma_i^z =\ket{\downarrow}$ the bit
state $x_i=0$. The mapping is such that the bit assignment that corresponds to the ground-state spin configuration of $H$ encodes
the solution of the computational problem. (Here we will mainly concern ourselves with a class of
problems that have a single solution, such that the ground state of
$H$ is non-degenerate.)

In the QAA algorithm, the classical spins (bits) are represented by quantum spin-1/2
degrees of freedom. The QAA protocol is typically carried out at zero
temperature and proceeds by preparing 
the system in a uniform superposition of all $\sigma_i^z$ eigenstates by applying a strong transverse field ($\sigma_i^z$ is a Pauli matrix that defines a local
quantization axis for the $i$-th spin). Annealing is implemented by
adiabatically ``turning off'' the transverse field while ``turning
on'' the Hamiltonian $H$. This process defines a time-dependent
Hamiltonian
\be
H_a = (1-s) H - s V\,,
\ee
where $V = \sum_{i=1}^N \sigma_i^x$ is the transverse field term and
$s=f(t)$ is a time-dependent parameter. The protocol usually starts
with $f(0)=1$ to ensure that the system is in the transverse-field
ground state, where each spin is polarized along the $x$-axis, and
ends with $f(T_f)=0$, which recovers the target Hamiltonian at time
$T_f$. The quantum adiabatic theorem~\cite{Qad1,Qad2} guarantees that
the system remains in its instantaneous ground state if $f(t)$ varies
``slowly enough'' with time. To be more specific, it says that the
total duration $T_f$ of the protocol should satisfy
\be
T_f \gg \hbar \frac{\text{max}_s|V_{10}(s)|} {(\Delta E_{\text{min}})^2} \,,
\ee
where $V_{m0}=\braket{0|\partial H_a/\partial s|m}$ in the eigenbasis
spanned by $\ket{m}$, $m=1,\dots,2^N-1$, and $\Delta E_{\text{min}}$
is the minimum gap between the ground state and the first excited
state encountered during the entire protocol.

Generally, $V_{m0}$ is proportional to system size for a local
Hamiltonian and the scaling of $T_f$ is controlled by the scaling of
the minimum gap. This implies that if the system passes through a phase 
transition where the gap vanishes, the time
to solution using a quantum annealing protocol can be polynomial or
exponential in system size depending on the behavior of the minimum gap with 
system size. Continuous phase transitions have a scale-invariant critical
point, which implies that the gap must have a polynomial dependence on
system size~\cite{SQPT}. First-order transitions, on the other hand,
manifest themselves in finite-size systems via gaps that vanish
exponentially with system size, although there are pathological cases
where the gap closing is only polynomial~\cite{FQPTgap}. It is
therefore highly probable that the QAA algorithm fails to find the
solution when the annealing protocol described by $H_a$ leads through
a first-order transition in the thermodynamic limit.

%%%%%%%%%%%%%%%%%%%%%%%%%%%%%%%%%%%%%%%%%%%%%%%%%%
\begin{figure}[t]
\includegraphics[width=\columnwidth]{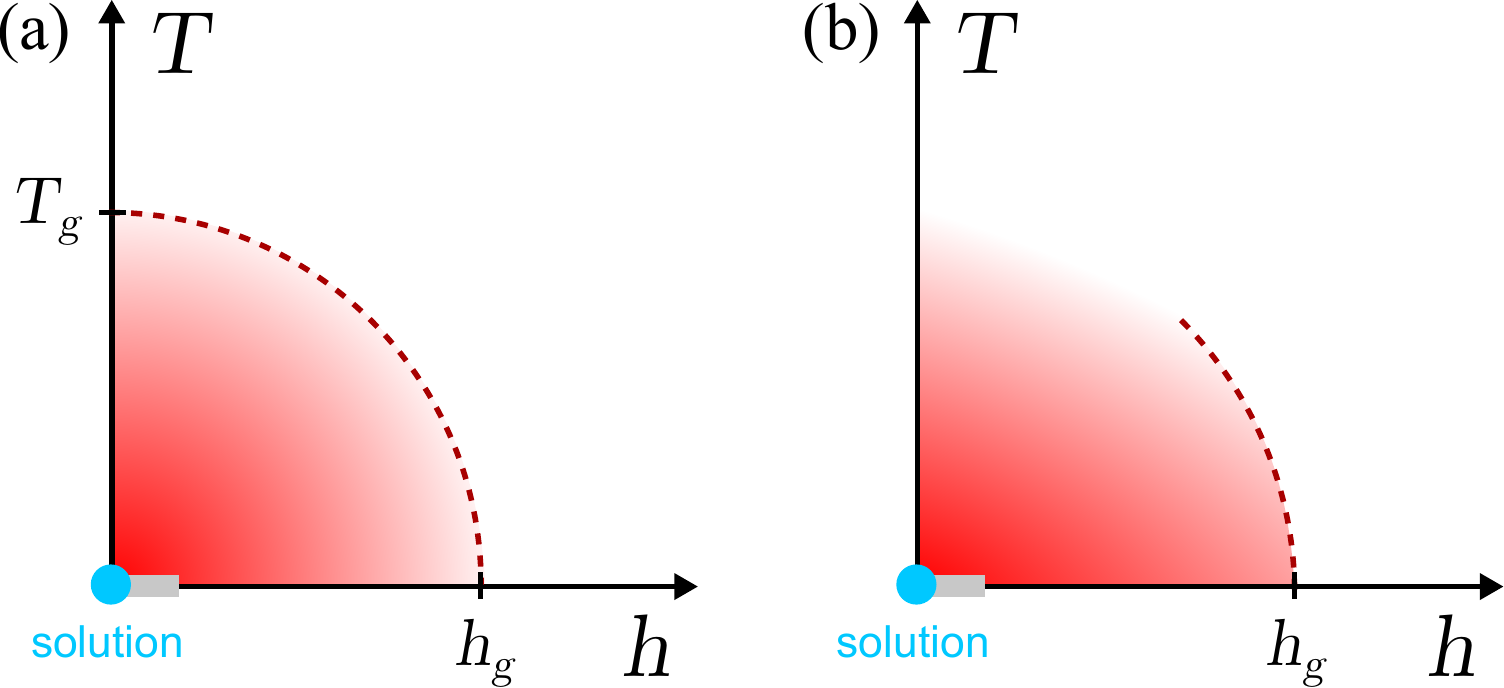}
\caption{Cartoon phase diagrams in the temperature-transverse field
  parameter space for two scenarios that may occur in the application
  of the SA and QAA algorithms to the solution of computational
  problems. In the scenario of panel (a), SA meets a transition to a
  glassy phase at $T=T_g$, whereas QAA encounters a first-order
  quantum phase transition at $h=h_g$. An example of this scenario is
  the 3-XORSAT problem, as it was formulated and studied in, e.g.,
  Refs.~\onlinecite{Ricci-Tersenghi2001,EF}. Dashed line indicates a
  putative phase boundary that terminates at the two critical points
  on the axes. Panel (b) depicts the scenario we introduce and study
  in this work, namely, the case where there is no bulk classical thermodynamic
  phase transition to a glass phase, but the obstruction of a
  first-order quantum phase transition nevertheless remains. An
  example of this scenario is the lattice embedding of 3-regular
  3-XORSAT we introduce in Sec.~\ref{sec:3xor}. The dashed line
  indicates a putative phase boundary that terminates at the quantum
  critical point, but does not extend all the way to the classical
  axis. In both panels, a thick grey line close to the origin
  delineates the range where the perturbative gap collapse argument of
  Ref.~\onlinecite{Alt} is potentially relevant and may obstruct the QAA
  protocol.}
\label{fig:cartoon}
\end{figure}
%%%%%%%%%%%%%%%%%%%%%%%%%%%%%%%%%%%%%%%%%%%%%%%%%%

In what follows, we will rewrite $H_a$ as
\be
H_a = J H - h V \,, \label{eq:Hanneal}
\ee
allowing $J$ and $h$ to take arbitrary positive values, to conform
with common notation in the literature.

Since $H$ is classical, one may also consider using simulated
annealing (SA) to reach its ground state. In this protocol, $h=0$ and
one slowly varies the temperature $T$ from $T=\infty$ to $T=0$. Local
thermal dynamics, implemented via, e.g., the Metropolis algorithm,
progressively lead toward lower-energy configurations. SA, and in fact
any local classical algorithm, fails whenever a first-order transition
into a glass phase is encountered upon reducing the temperature. This
is true regardless of the hardness of the computational problem
encoded by $H$~\cite{Ricci-Tersenghi2001,Ricci-Tersenghi2010}.

When SA and QAA are taken on equal footing as methods for the solution
of a given problem, they give rise to a phase diagram as a function of
$T$ and $h$, whose origin represents the solution of the
problem. Fig.~\ref{fig:cartoon} shows two distinct scenarios for
this phase diagram. In the first scenario, the solution is separated
from both the classical high-$T$ paramagnet and the strong-field
quantum paramagnet, i.e., the initial states of the SA and QAA
protocols, respectively, by first-order transitions. This is a
commonly encountered scenario for computational problems, such as
satisfiability or coloring, and is illustrated in
Fig.~\ref{fig:cartoon}(a). For example, the 3-regular 3-XORSAT problem,
as it was formulated and studied in Ref.~\onlinecite{EF} and also briefly
introduced below, belongs to this category.

In this work, we will use a lattice reformulation of computational
problems that lacks the classical transition to a glassy
phase. This formulation introduces an alternative scenario to the
aforementioned one and raises the question of whether the quantum
phase transition is absent as well in this case, i.e., whether the two
transitions are somehow linked. Below we will provide evidence for a
negative answer to this question: our results suggest that the quantum
phase transition remains present and first-order, and hence most
probably accompanied by exponentially vanishing gaps in progressively
larger finite-size systems, even in the absence of a thermodynamic
classical transition to a glassy phase.

%%%%%%%%%%%%%%%%%%%%%%%%%%%%%%%%%%%%%%%%%%%%%%%%%%%%%%%%%%%%%%%%%%%%%%%%%%%%%
\section{$k$-regular $k$-XORSAT and lattice embedding}\label{sec:3xor}

\subsection{The XORSAT problem}\label{sec:originaldef}

In this section, we describe the mapping of the the $k$-regular
$k$-XORSAT problem~\cite{Xref} to a spin Hamiltonian. We choose XORSAT
because it is a prototypical problem in both physics and theoretical
computer science. Even though XORSAT can be solved in polynomial time
with Gaussian elimination, it nevertheless has evaded efficient
solution with any local algorithm, including variants of the
Davis-Putnam algorithm~\cite{Haanp2006}, message-passing
methods~\cite{Jia2005}, stochastic search~\cite{XNL2}, simulated
annealing~\cite{Ricci-Tersenghi2001}, and quantum adiabatic
annealing~\cite{EF}.

Here we focus on the $k$-regular variant of $k$-XORSAT. This
constraint satisfaction problem is defined on $N$ Boolean variables
subject to $N$ clauses, where each clause takes in $k$ bits and each
bit participates in $k$ clauses. The solution to the problem is a bit
assignment that satisfies all clauses. An XORSAT clause evaluates to 0
(false) or 1 (true) if the sum of the bits in the clause modulo 2 is 0
or 1, respectively. In spin language, this can be interpreted as
requiring the product of the spins in a particular clause to be
positive or negative and associating an energy cost to the unfavorable
outcome. For example, the spin Hamiltonian can be written as
\be
\label{rh}
H = - \sum_{j=1}^N \prod_{i \in c_j} \sigma^z_i \,,
\ee
where $c_j$ is the set of the $k$ indices of the spins that
participate in the $j$-th clause, for $k$ odd. Since clauses are not
constrained to connect nearby spins only, this spin Hamiltonian is
best represented as a bipartite $k$-regular graph, where one
independent set of vertices represents the spins and the other the
clauses. A random instance of this problem is thus a randomly
generated bipartite $k$-regular graph. A solution of an instance (if
it exists) is given by a corresponding ground state of $H$.

Numerical examination of the QAA algorithm for Hamiltonian~\eqref{rh}
restricted to $k=3$ and to instances with unique ground states showed
that the minimum gap closes exponentially with system size, indicating
a first-order transition at a particular value of the transverse field
in the thermodynamic limit~\cite{EF}. This finding implies that QAA
takes an exponentially long time to find the solution in this
formulation of 3-regular 3-XORSAT. On the other hand, application of
the SA algorithm to the XORSAT problem reveals a random first-order
transition into a glassy phase at some characteristic temperature ---
see, e.g., Ref.~\onlinecite{Ricci-Tersenghi2001}. These results suggest that
the solutions of XORSAT instances reside deep inside a glass phase and
are inaccessible to both classical local search algorithms and QAA,
despite the fact that XORSAT is computationally tractable (i.e., in
complexity class P).

\subsection{Lattice embedding}\label{sec:latrep}

In an attempt to avoid the aforementioned obstructions to the solution
of XORSAT, here we introduce a lattice embedding of the problem that
circumvents the classical thermodynamic transition. The idea is based on
previous works by some of us~\cite{Chamon2016,Lei}. Note that this
lattice embedding does not enable an efficient solution of the problem
via SA, despite the absence of the glass transition, as the dynamics
instead becomes glassy upon approaching $T=0$. The rationale for this
reformulation is rather to see whether the avoidance of the
thermodynamic glass transition and hence the absence of a finite-$T$
classical glass phase has any effect on the quantum axis.

\begin{figure}[t]
\includegraphics[width=\hsize]{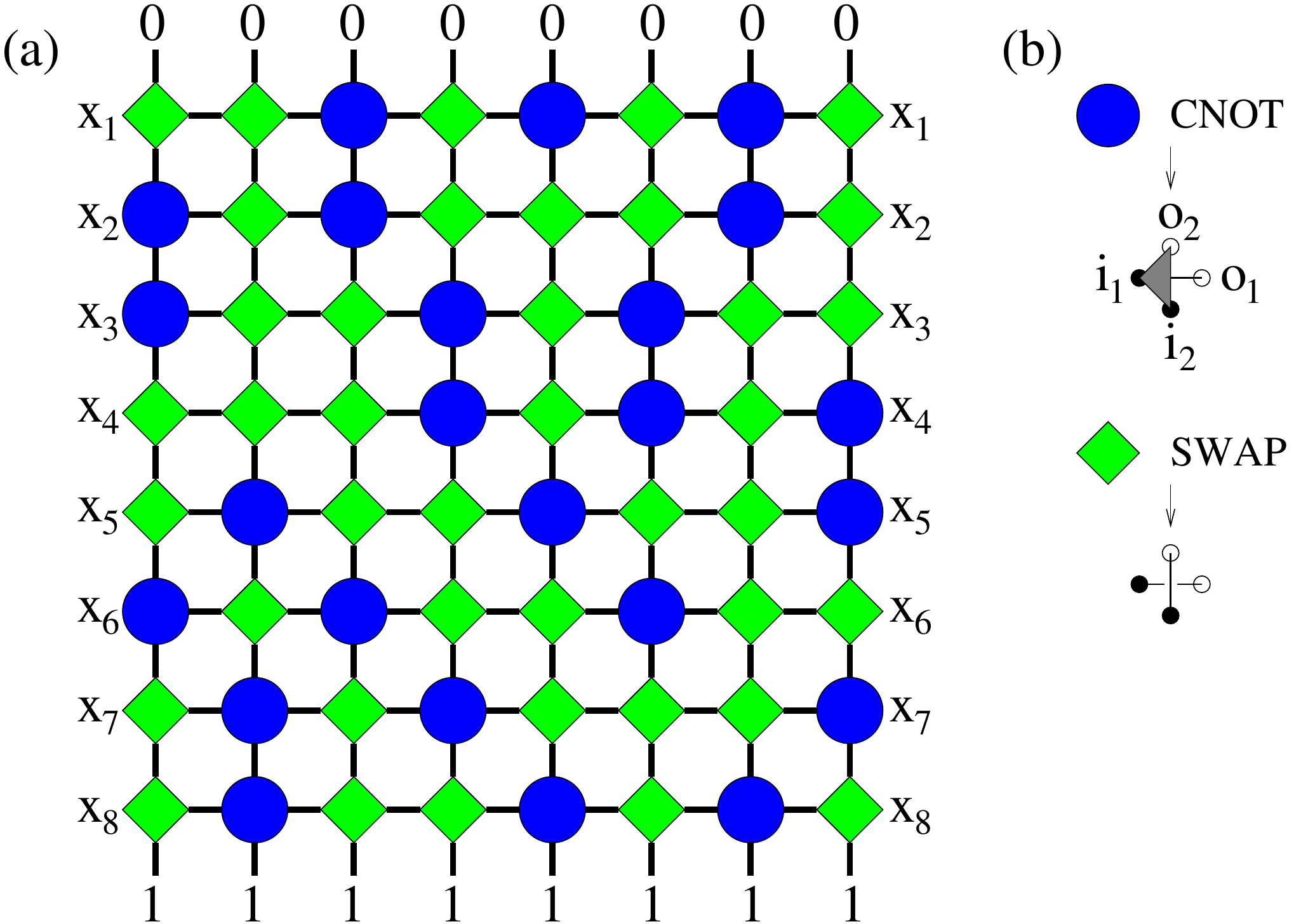}
\caption{(a) Lattice representation of 3-regular 3-XORSAT instance
  with 8 variables and 8 clauses as a $8\times 8$ lattice of CNOT and SWAP
  gates. There are 4 bits coupled by a gate at each position of the
  lattice, as shown in (b).  The variable bits $x_i$ record the
  solution of the problem upon termination of a protocol that reaches
  the ground state. Upper and lower boundary states are forced by a
  strong field that favors the uniform bit states shown. (b) Sketch of
  couplings between bits in each of the gates. Lines denote
  ferromagnetic bonds in the spin representation and grey triangle
  represents the 3-spin term in Eq.~\eqref{eq:Cn}. The formula for the
  3-XORSAT instance shown here is $(x_2 \oplus x_3 \oplus x_6) \wedge
(x_5 \oplus x_7 \oplus x_8) \wedge
(x_1 \oplus x_2 \oplus x_6) \wedge
(x_3 \oplus x_4 \oplus x_7) \wedge
(x_1 \oplus x_5 \oplus x_8) \wedge
(x_3 \oplus x_4 \oplus x_6) \wedge
(x_1 \oplus x_2 \oplus x_8) \wedge
(x_4 \oplus x_5 \oplus x_7)$.}
\label{fig:lat}
\end{figure}

The lattice embedding is achieved by drawing each variable and each
clause as a bit line or ``bus'' and laying all lines on a 2D plane,
with vertical lines corresponding to clauses and horizontal ones to
variables, as shown in Fig.~\ref{fig:lat}. Each variable (clause)
corresponds to a horizontal (vertical) Ising spin chain and the
intersection between variable chains and clause chains is mediated by
CNOT or SWAP gates. If the $y$-th variable participates in the $x$-th
clause, then a CNOT gate is placed at the intersection of the $x$-th
vertical line with the $y$-th horizontal line, else a SWAP gate is
placed to ensure that the bit and clause do not couple.
%For example, the realization of 3-regular-3-XORSAT shown in Fig.~\ref{fig:lat}
%has 8 bits and 8 clauses which can be written as
%\begin{align}
%Y_1=(x_2 \oplus x_3 \oplus x_6),\\
%Y_2=(x_5 \oplus x_7 \oplus x_8),\\
%Y_3=(x_1 \oplus x_2 \oplus x_6),\\
%Y_4=(x_3 \oplus x_4 \oplus x_7),\\
%Y_5=(x_1 \oplus x_5 \oplus x_8),\\
%Y_6=(x_3 \oplus x_4 \oplus x_6),\\
%Y_7=(x_1 \oplus x_2 \oplus x_8),\\
%Y_8=(x_4 \oplus x_5 \oplus x_7).
%\end{align}
%A satisfying assignment $\{x_1,x_2,...,x_8\}$ would yeild $Y_i=1$ for all $i$,
%as required by the bottom boundary in Fig.~\ref{fig:lat}.

Each gate has two inputs ($i_1$ and $i_2$) and two outputs ($o_1$ and
$o_2$).  The gate constraints can be written in spin language as
\begin{subequations}
\label{eq:Hxorlatt}
\begin{equation}\label{eq:Sw}
H^{0}_{x,y} = - \sigma^z_{x,y;i_1} \sigma^z_{x,y;o_1} - \sigma^z_{x,y;i_2}
\sigma^z_{x,y;o_2}
\end{equation}
for a SWAP gate and
\begin{equation}\label{eq:Cn}
H^{1}_{x,y} = - \sigma^z_{x,y;i_1} \sigma^z_{x,y;o_1} + \sigma^z_{x,y;i_1}
\sigma^z_{x,y;i_2} \sigma^z_{x,y;o_2}
\end{equation}
for a CNOT, where the subscripts $x$ and $y$ specify the position of
each gate. To accommodate the $y$-th variable to appear negated in the
$x$-th clause, we can simply change the sign of the second term
in~\eqref{eq:Cn}. Here we will deal only with monotone instances where
no variables appear negated, and hence no such change will be
necessary.

Inter-site ferromagnetic bonds of strength $J$ are placed between the
outputs of a gate and the inputs of nearest-neighboring gates. This
construction ensures that when all ferromagnetic bonds are satisfied
by a spin configuration, the corresponding bit assignment satisfies
all clauses and is the solution to the problem. The overall spin
Hamiltonian is
\begin{align}
\label{eq:H}
H =&{\ } - \sum_{\braket{x,y;x',y'}} \sum_{v=1,2} \sigma^z_{x,y;o_v} \sigma^z_{x',y';i_v}
         + g \sum_{x,y} H_{x,y}^{A_{x,y}} \\
\label{eq:H1}
   &{\ } - \sum_x (\sigma^z_{x,0;o_2} \sigma^z_{x,\partial_{y=0}} 
         + \sigma^z_{x,N;i_2} \sigma^z_{x,\partial_{y=N}})\\
\label{eq:H2}
   &{\ } + g \sum_x (\sigma^z_{x,\partial_{y=0}} \pm \sigma^z_{x,\partial_{y=N}}) \,.
\end{align}
\end{subequations}
The sums are over $x,y = 1,\dots,N$, so that a problem instance with
$N$ variables maps to a lattice with $N^2$ sites with a gate at each
site, $\braket{\dots}$ denotes neighboring positions on the lattice,
$v$ is the orientation of the bond labeled as 1 (2) for horizontal
(vertical), and $g$ is a constant that offsets the energy cost of the
gate and boundary terms with respect to the bond terms.

%%%%%%%%%%%%%%%%%%%%%%%%%%%%%%%%%%%%%%%%%%%%%%%%%%%%%%%%%%%%%%%%%%%%%%%%%%%%%%
%%%%%%%%%%%%%%%%%%%%%%%%%%%%%%%%%%%%%%%%%%%%%%%%%%%%%%%%%%%%%%%%%%%%%%%%%%%%%%

The first line of Eq.~\eqref{eq:H} defines the interactions in the
bulk.  $A$ is the biadjacency matrix of the bipartite $k$-regular
graph that defines the problem instance, as described above. When
$A_{x,y}=1$, a CNOT gate is placed at position $(x,y)$, otherwise a
SWAP is placed there instead, as sketched in Fig.~\ref{fig:lat}. Note
that in our convention indexing proceeds from left to right and top to
bottom.  We consider the limit $g \to \infty$, so that outputs are
essentially ``dummy'' spins, whose state is completely controlled by
the gate inputs. When the ground state of $H$ is reached, the solution
appears on the left and right boundaries of the lattice, which are
left free.  The requirement that clauses sum modulo 2 to 0 or 1 is
enforced by the term on the last line, which acts only on the top and
bottom rows of boundary spins and whose relative sign between top and
bottom depends on whether $k$ is odd or even. For example, for $k=2$
we choose $+$ (all-zeros state in bottom row) to ensure that all
clauses sum modulo 2 to 0, whereas for $k=3$ we choose $-$ (all-ones
state in bottom row), which requires all clauses to sum modulo 2 to 1.
When $g\to\infty$ this interaction becomes a hard constraint. This
clause constraint is then propagated to the bulk by the terms in
Eqs.~\eqref{eq:H1} and ~\eqref{eq:H2}, which define the interaction
between spins at the top ($\partial_{y=0}$) and bottom
($\partial_{y=N}$) boundaries and bulk gate spins.

Here we restrict our analysis to instances of $k$-regular $k$-XORSAT
which have all the minimum nonzero number of solutions. For $k=3$,
generic problems with a unique solution exist. For all such problems,
a spin reversal transformation exists which maps the solution to the
all spins down state~\cite{EF} and we shall assume that our system
has already undergone this transformation. We will focus on these
problems below, as they are a finite fraction of all 3-XORSAT
instances and are thus good representatives of the full
ensemble~\cite{Jorg2010}. For $k=2$, there are always two solutions,
one of which corresponds to the all-down state.

We generate 3-regular 3-XORSAT instances with unique solutions by
first generating a random bipartite 3-regular graph and retaining only
those instances which have an odd determinant, as this condition
enforces a unique solution~\cite{GE}. We also ensure that the
generated graph is connected. We then use the biadjacency matrix of
this graph to define the lattice embedding. For the $k$-regular
variants of XORSAT, the finite lattices are by definition
square. Varying the clause-to-variable ratio amounts simply to
changing the lattice aspect ratio.

%%%%%%%%%%%%%%%%%%%%%%%%%%%%%%%%%%%%%%%%%%%%%%%%%%%%%%%%%%%%%%%%%%%%%%%%%%%%%
\section{Analytic results: limiting cases and weak-field perturbation theory}
\label{sec:anres}

\subsection{Dilute constraint limit and 2-XORSAT}

In the lattice setup, if we only have SWAP gates at all the
intersections between clauses and bits, then we recreate disconnected
transverse field Ising chains. This can be seen by considering the
action of the SWAP gate as given in Eq.~\eqref{eq:Sw}, where we see
that spins are only coupled along either the horizontal or vertical
directions. Considering also the transverse-field term, and
remembering that input and output spins of gates are locked when
$g\to\infty$, the Hamiltonian for a decoupled chain reduces to
\begin{equation}
\label{eqHI}
H_a = - J \sum_{i=1}^N \sigma^z_i \sigma^z_{i+1} - h \sum_{i=1}^N \sigma^x_i \,,
\end{equation}
where $i$ now denotes the coordinate along the chain. This is simply
the transverse-field Ising (TFI) chain, which is known to have a
continuous phase transition at $h=J$ and a critical behavior which is
well understood \cite{PTFI}. This would then imply that in the limit
of zero CNOT gates on the lattice, we would have a second-order
transition characterized by the TFI chain universality class.

The lattices that arise in our embedding of $k$-regular $k$-XORSAT
have $k$ CNOTs in each vertical and horizontal line. This leads to a
CNOT density $\frac{k}{N}$. For $N \gg k$, the system reduces to
independent TFI chains coupled at a vanishing number of points. Two
possibilities arise for a potential phase transition that the system
may undergo as a function of $h$. The first is that the phase
transition remains continuous as for decoupled TFI chains when the
density of ``impurities'' is vanishing. The second possibility is that
this vanishing number of impurities drastically changes the nature of
the phase transition from continuous to first-order. In the first
case, we would be left with a lattice which is able to solve the
computational problem in polynomial time. In the second case, the
lattice would require exponential time and would be an example of a
system where adding a vanishing number of impurities changes the order
of the transition. An example of this behavior occurs in the
polymerization of rubber~\cite{VR}, where the process of vulcanization
leads to a vanishing number of cross-links between polymers, which in
turn changes the state of rubber from liquid to solid.

\begin{figure}[t]
\includegraphics[width=0.85\hsize]{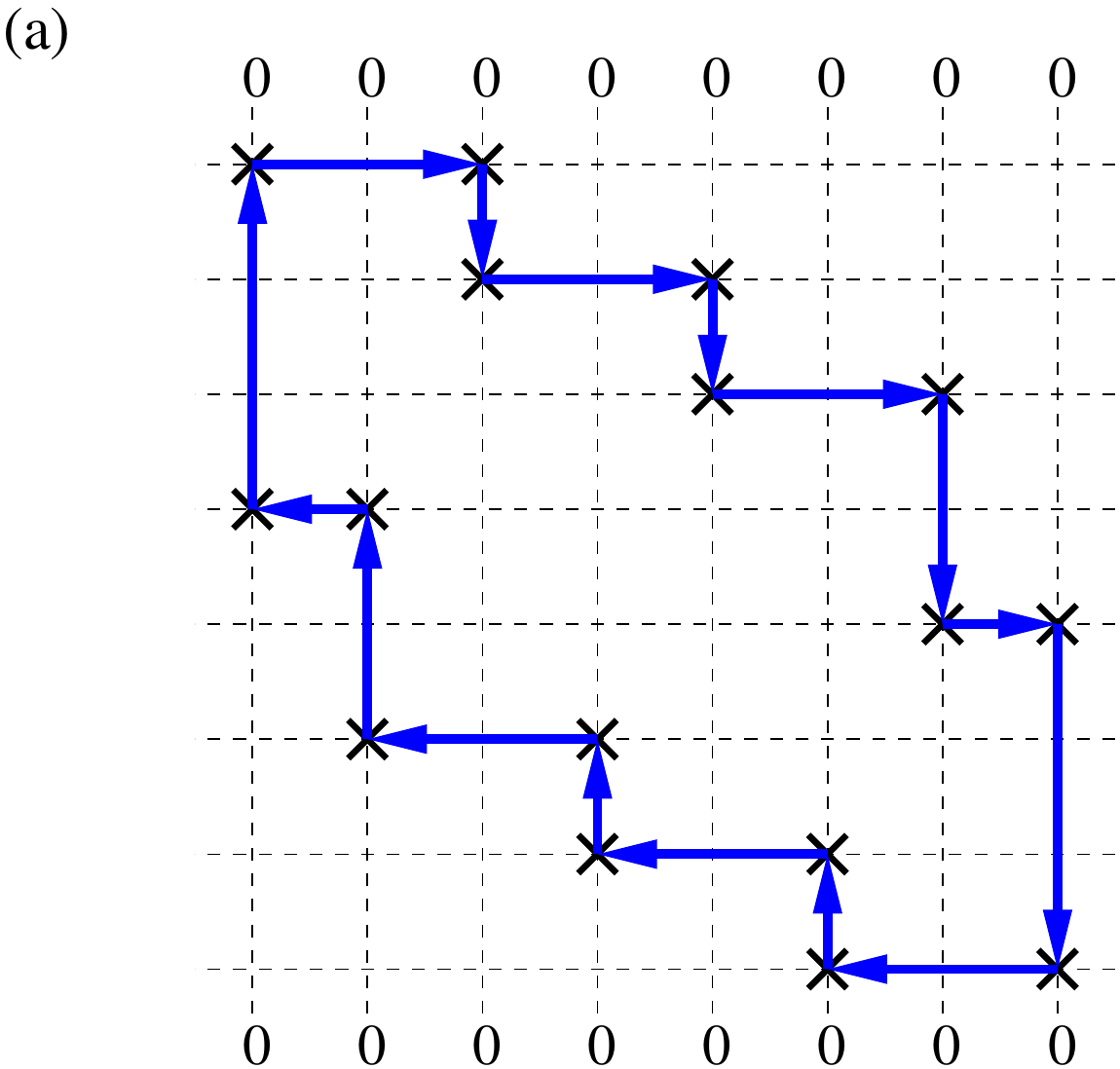}\vspace{2em}
\includegraphics[width=0.9\hsize]{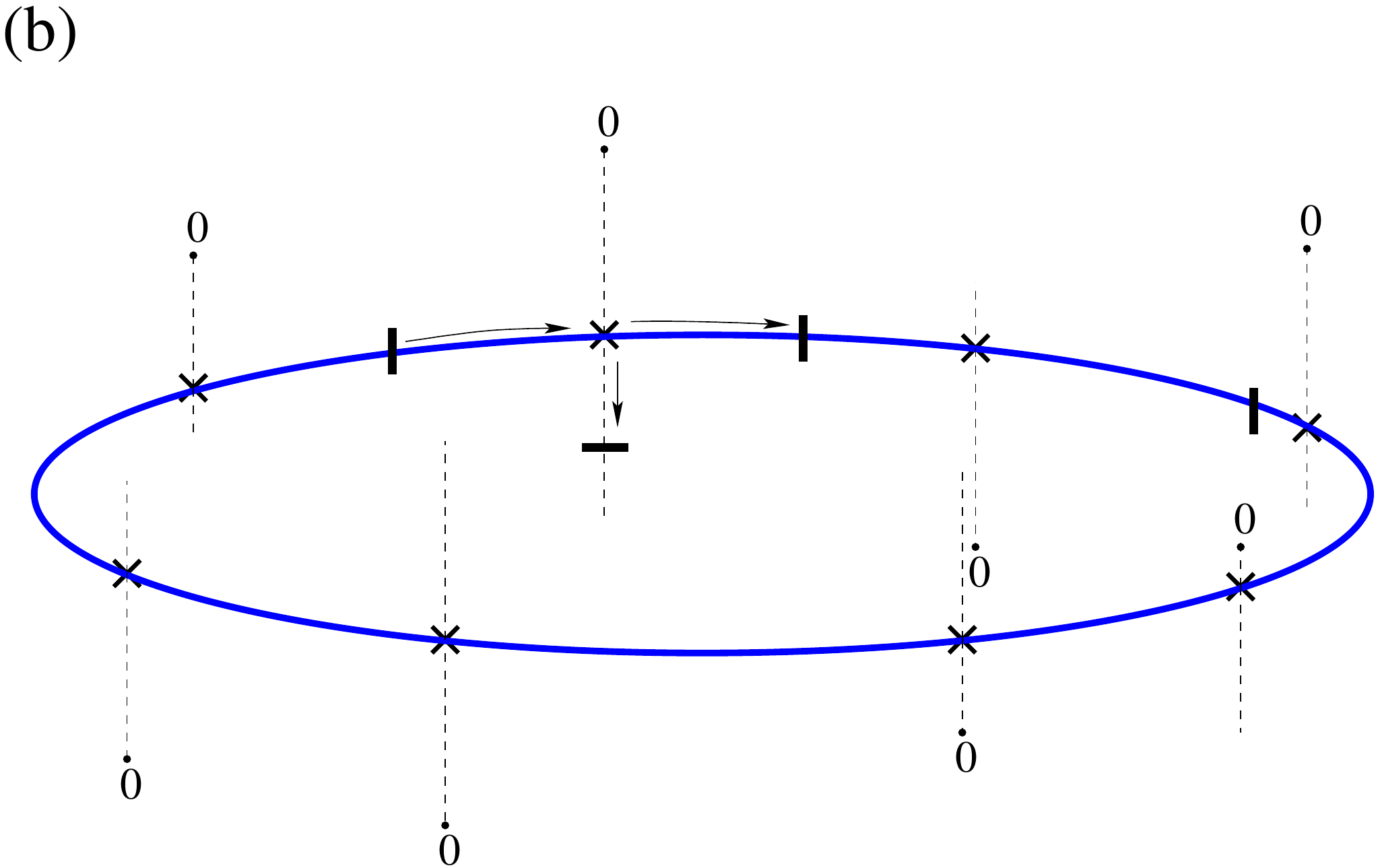}
\caption{(a) ``Backbone'' loop for a $k = 2$ instance of $k$-regular
  $k$-XORSAT in the lattice representation. SWAP gates are omitted and
  CNOT gates are denoted by $\times$ symbols. Arrows indicate the
  ``unraveling'' procedure described in the text. (b) The
  corresponding unraveled loop picture with spin chains radiating out
  of CNOT gates. Domain walls are sketched as $\boldsymbol|$ symbols
  and their movement, indicated by the arrows, heals broken bonds and
  reduces energy.}
\label{fig:dw}
\end{figure}

Let us examine the $k=2$ case. Each instance contains a periodic Ising
chain, as it corresponds to a series of ferromagnetic bonds between
spins, where each spin participates in only two bonds, as illustrated
in Fig.~\ref{fig:dw}(a). This lattice can be reconfigured as an Ising
chain with offshoots, as shown in Fig.~\ref{fig:dw}(b), using the
following ``unraveling'' procedure. First, pick an arbitrary CNOT and
an arbitrary direction (vertical or horizontal), then draw a link
between the starting CNOT and its neighbor in that direction. As there
is only one neighbor in either direction, there is no ambiguity in
this step. Now rename the neighbour as the starting site and follow
the same procedure using the direction perpendicular to the current
direction. This process creates a unique loop with spin chains
branching out at the locations with CNOTs. This equivalence is valid
as SWAP gates only braid chains over each other without
interactions. Taking the limit of large size, we would expect the
average separations between CNOTs to be of order $N$ and the
fluctuations about this should be statistically small.

Examining the energetics of domain walls in this system illustrates
why SA is expected to be efficient in reaching the solution in this
case. Let us consider a configuration of this system with a number of
domain walls which would correspond to a typical state encountered at
finite temperature, as seen in Fig.~\ref{fig:dw}(b). If we translate a
domain wall through a CNOT, it generates two domain walls on the other
side, one of which can be healed by translating it out to the
boundary, while the other can travel around the ring until it meets
another domain wall, with which it can mutually annihilate. In this
way, domain walls can be healed all the way to a state without domain
walls, i.e. the ground state, in a smooth sequence of steps that
monotonically reduce energy other than the one additional bond that
must be broken when passing through a CNOT. In Sec.~\ref{sec:QMCres},
we will show that 2-XORSAT is also efficiently solved with QAA.

%%%%%%%%%%%%%%%%%%%%%%%%%%%%%%%%%%%%%%%%%%%%%%%%%%%%%%%%%%%%%%%%%%%%%%%%%%%%%%
%%%%%%%%%%%%%%%%%%%%%%%%%%%%%%%%%%%%%%%%%%%%%%%%%%%%%%%%%%%%%%%%%%%%%%%%%%%%%%

%%%%%%%%%%%%%%%%%%%%%%%%%%%%%%%%%%%%%%%%%%%%%%%%%%%%%%%%%%%%%%%%%%%%%%%%%%%%%%
\begin{figure}[t]
\includegraphics[width=0.85\hsize]{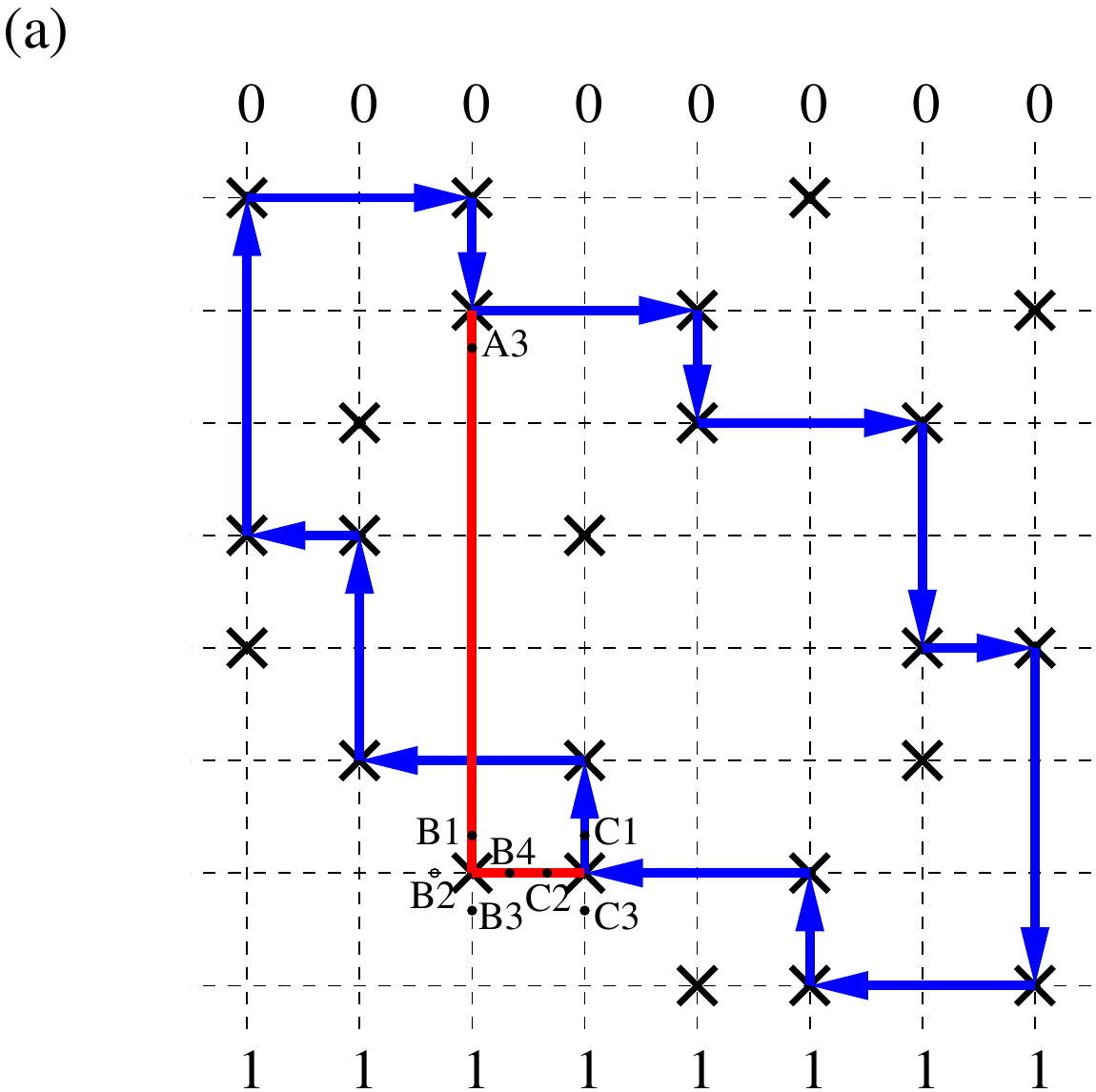}\vspace{2em}
\includegraphics[width=0.9\hsize]{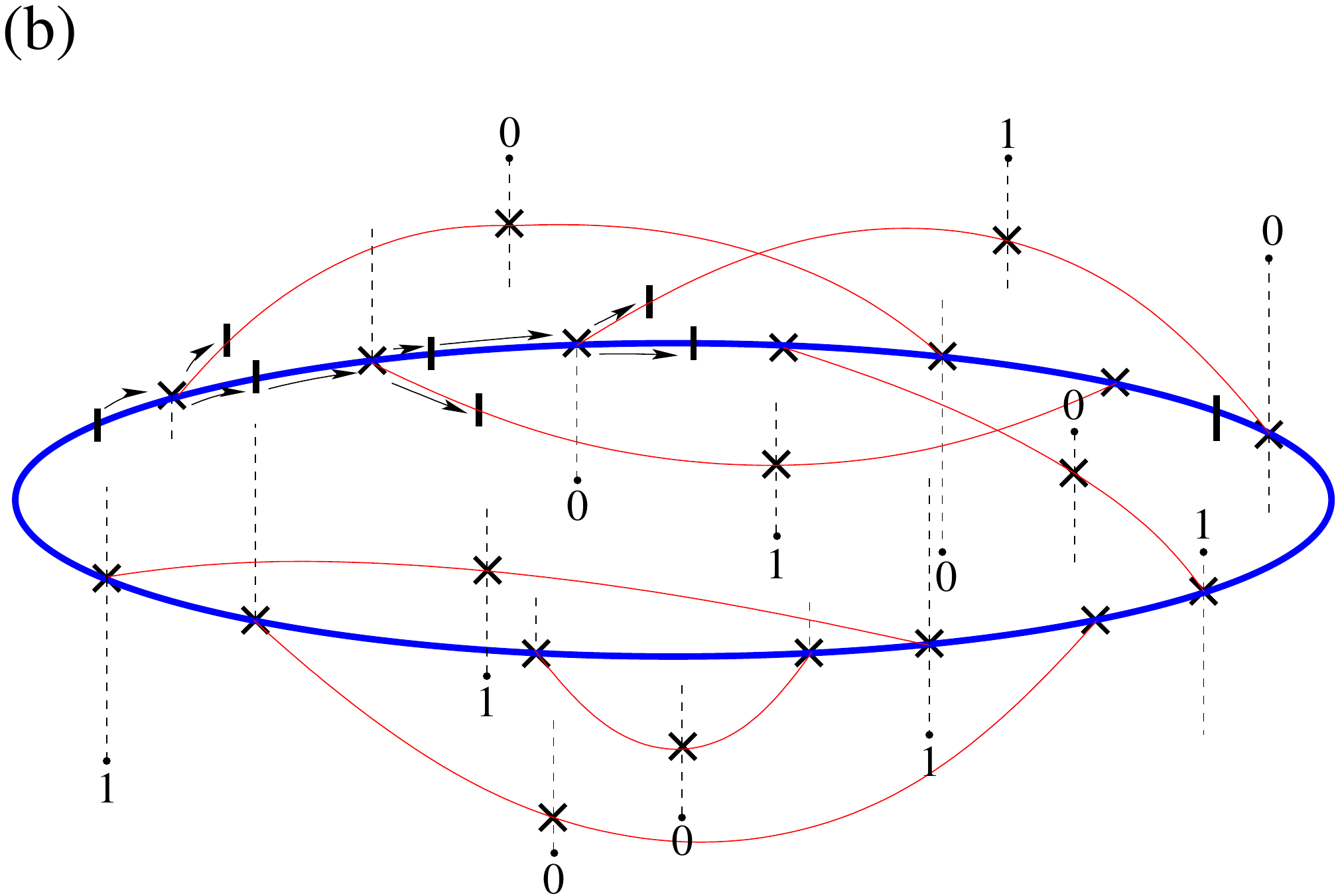}
\caption{(a) Lattice representation of 3-regular 3-XORSAT instance
  with the same backbone as the one shown in Fig.~\ref{fig:dw}(a). The
  backbone cannot independently fluctuate between positive and
  negative values, due to constraining couplings to spins outside the
  backbone (example shown in red), and hence the backbone cannot be
  isolated as in the $k=2$ case. (b) Loop equivalent for the $k=3$
  realization, with arrows showing domain wall movement that in this
  case involves branching of domain walls at CNOT gates.}
\label{fig:3XL}
\end{figure}
%%%%%%%%%%%%%%%%%%%%%%%%%%%%%%%%%%%%%%%%%%%%%%%%%%%%%%%%%%%%%%%%%%%%%%%%%%%%%%

We now apply the same argumentation to the lattice representation of
the $k=3$ case. We decompose the lattice into a loop using the
unraveling procedure. Fig.~\ref{fig:3XL} shows a realization of
3-regular 3-XORSAT with the same backbone structure as
Fig.~\ref{fig:dw}. The additional CNOT gates now provide frustrating
interactions which force the backbone chain to have all spins pointing
down and spin inversion symmetry for that section of the lattice is
lost. We can look at this in detail using the particular
cross-connecting CNOT shown in Fig.~\ref{fig:3XL} (highlighted in red)
for the classical ground state where all bonds must be satisfied. If
we assume that the spins lying in the blue chain are +1, then spin A3
would be forced to -1, implying B1 has to be -1. From this it follows
that B2 must be +1, as B3 is +1 due to a direct connection to the
boundary. This would then force B4 and as a result C2 to be +1. And as
C3 is +1 due to the boundary, C1 must both be -1 creating a
contradiction as C1 belongs in the blue chain and must be +1. In terms
of the ring structure in Fig.~\ref{fig:3XL}(b), this would mean
cross-connections between various offshoots, which would destroy the
one-dimensional nature of the chain.

The convoluted loop structure of $k=3$ instances implies that domain
wall movement now becomes highly non-trivial: we cannot simply heal
domain walls by moving them to the boundary, but must instead
translate them to the next CNOT, where they can perhaps annihilate by
merging with another domain wall. However, moving a domain wall around
the ring now produces a large number of domain walls, as each CNOT
results in branching. Each of the resulting defects can be healed only
after traversing half the ring on average. This illustrates why SA
will fail to solve this problem efficiently, even though our lattice
formulation can be shown to feature no thermodynamic glass
transition~\cite{Chamon2016,Lei} and is also found to be similar to
the vertex models proposed in Ref.~\onlinecite{Chamon2016} (see also
Appendix~\ref{sec:app}).

%%%%%%%%%%%%%%%%%%%%%%%%%%%%%%%%%%%%%%%%%%%%%%%%%%%%%%%%%%%%%%%%%%%%%%%%%%%%%
\subsection{Weak-field perturbation theory: absence of gap collapse}\label{sec:PT}

%%%%%%%%%%%%%%%%%%%%%%%%%%%%%%%%%%%%%%%%%%%%%%%%%%%%%%%%%%%%%%%%%%%%%%%%%%%%%%
\begin{figure}[t]
\includegraphics[width=0.8\hsize]{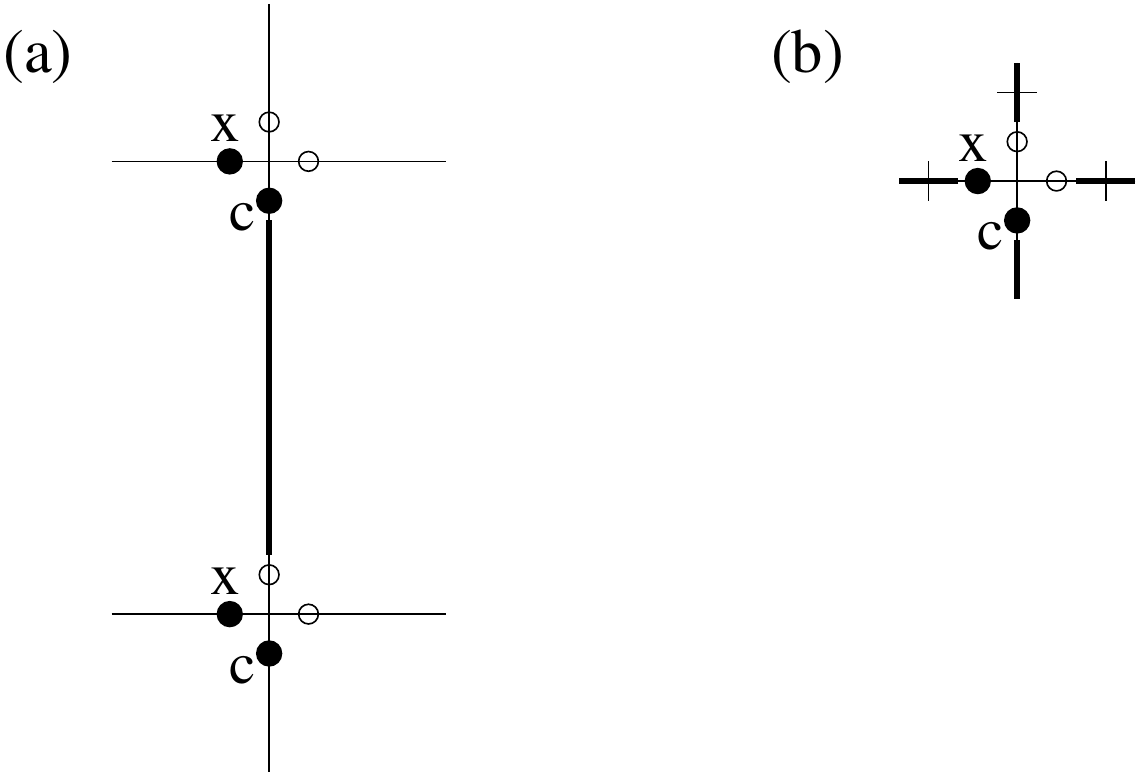}
\caption{(a) Sketch of $y$-bond connecting the spins of two gates.
  The state of this bond only affects the four perturbation correction
  terms corresponding to the spins shown as filled circles. (b)
  Flipping the control spin (X) of a CNOT gate switches the state of
  the bonds marked with a cut. Along the vertical line, either the
  upper or lower bond can be broken (upper bond shown here).}
\label{fig:XC}
\end{figure}
%%%%%%%%%%%%%%%%%%%%%%%%%%%%%%%%%%%%%%%%%%%%%%%%%%%%%%%%%%%%%%%%%%%%%%%%%%%%%%

Altshuler \emph{et al.}~\cite{Alt} pinpointed a potential setback
inherent in the QAA protocol. Perturbative analysis showed that
generic classical Hamiltonians set up to solve computational problems
may give rise to an avoided level crossing when an arbitrarily small
transverse field is added. This implies an exponential reduction in
annealing velocity, in order to maintain fidelity with the target
ground state. In order to fully address the potential issue of avoided
level crossings and to show that it is not present in our lattice
formulation, we will now present a full perturbative treatment of the
general $k$-regular $k$-XORSAT problem. A similar analysis appears in
Ref.~\onlinecite{Knysh2010}.

Consider the spectrum of a classical spin model that represents a
computational problem. We again restrict ourselves to instances with a
unique solution. The ground state of the model is therefore
unique. The first excited state is degenerate and is made out of all
those configurations that differ from the ground state in the status
of a single bond. At $h=0$, the corresponding energy levels have a
difference of $J$ from the ground-state level. Altshuler \emph{et al.}
show that perturbative corrections to the gap between the ground state
and the first excited state for small $h$ should depend on system size
in the general case, an argument resulting from the disorder present
in most spin representations of computational problems. This leads to
a second-order correction proportional to $Nh^2$, where $N$ is the
system size. They conclude that for arbitrarily small $h$ there exists
$N$ such that the correction surpasses $J$, leading to an avoided
level crossing.

Contrary to the above argument, we will show that there are no
vanishing gaps in the weak-field limit for $k$-regular $k$-XORSAT.
First, let us define a bond to have two states, the set state where it
is satisfied and the broken state, and the two inputs of a gate to be
degrees of freedom (as mentioned in the previous section, the outputs
of the gates are completely determined by the inputs, so they are not
degrees of freedom, and thus will be ignored). Each gate has four
bonds radiating out. Flipping either of the two input spins of a SWAP
gate flips the states of two bonds, irrespective of what their states
are and irrespective of what the state of the spin was. For the CNOT
gate, flipping the carry input (C) switches the state of two bonds,
but flipping the control input (X) switches the state of three bonds,
again irrespective of the initial bit states
(Fig.~\ref{fig:XC}). Flipping a boundary spin switches the state of
only one bond. The important thing to note here is that the cost
induced due to a flip does not depend on the states of spins.
Therefore, perturbation theory on the ground state creates $N$
corrections, but none of these depend on the ground-state
configuration.

First, consider that all excited-state levels are non-degenerate. This
situation is artificial and we treat it only because it simplifies the
discussion of the physically relevant case of highly degenerate
excitation manifolds discussed below. The excitation energy
$E^{\mathrm{flip}}_i$ of the state created due to a single spin flip
at site $i$ is one of $J, 2J, 3J$, depending on whether it belongs to
the boundary, a SWAP gate or a CNOT gate
respectively. The second-order perturbation correction to the
ground-state energy is
\begin{equation}
\Delta E^{(0)} = \sum_{i=1}^N \frac{h^2}{E^{\mathrm{flip}}_i} \propto \frac{N h^2}{J} \,,
\end{equation}
so far in accordance with the Altshuler \emph{et al.} argument. Any
term in this sum is determined only by the state of the four bonds
radiating out of a particular point on the lattice and the gate
corresponding to that point. Each lattice point carries two spins,
implying that the status of a particular bond, which can only control
two lattice points, is relevant only to at most four terms in this
sum. As the only difference between the ground state and one of the
first excited states is the status of one bond, only four or lesser
number of terms in the perturbation corrections for both states can
differ. It follows that the change in the gap due to these corrections
is
\begin{equation}
|\Delta E^{(0)} - \Delta E^{(1)}| \propto h^2/J \,,
\end{equation}
which is \emph{independent} of system size, similar to the clean TFI
model~\cite{IMPT}.

We now take into account the mixing of the degenerate first-excited
states amongst themselves due to the transverse field. This typically
results in the degenerate levels spreading over a band of width
$\propto h$ and is, once again, independent of system size.
%An example of this can be seen for the transverse field Ising chain~\cite{IMPT}.
After first-order degenerate perturbation theory, the states that
result from the mixing of the first-excited configurations at zero
field are
\begin{equation}
\ket{\tilde E^{(l)}} = \sum_i \alpha^{(l)}_i \ket{E^{(1)}_i} \,,
\end{equation}
where $\ket{E^{(1)}_i}$ is a first-excited configuration with energy
$E^{(1)}$. The states $\ket{\tilde E^{(l)}}$ are then used to perform
higher-order perturbation theory. Second-order corrections look like
\begin{equation}
\Delta \tilde E^{(l)}= h^2 \sum_{i,j} \sum_m
\frac{\alpha^{(l)}_i \alpha^{(l)}_j}{|E^{(1)}-E_m|} \braket{E^{(1)}_i|V|m} \braket{m|V|E^{(1)}_j} \,,
\end{equation}
where $m$ runs over states that are necessarily not $\ket{E^{(1)}_i}$.
Of the ``diagonal'' terms $|\braket{E^{(1)}_i|V|m}|^2$, at most four
are nonzero, by the same reasoning we employed above. For the terms
with $i \neq j$, if $\ket{E^{(1)}_i}$ has a frustrated bond in the
bulk, there is no way to connect it to $\ket{E^{(1)}_j}$ via $V$ using
states $\ket{m}$ within the first level. On the other hand, when
$\ket{E^{(1)}_i}$ contains a frustrated boundary bond, we can connect
to all other such singly excited states. However, these second-order
processes are equally possible for all states, regardless of gate
configuration, and hence this correction amounts to a uniform shift of
all levels. The above means that the relative shift of a first-excited
state level with respect to the ground-state level is going to be
bounded in the same way as it was for just one spin configuration in
the first level. The same argumentation extends to higher corrections
straightforwardly. We therefore conclude that there is no perturbative
gap collapse in the weak-field limit for the models we study here.

We remark that the arguments presented above also extend to the random
graph spin model for 3-regular 3-XORSAT studied in Ref.~\onlinecite{EF}, and
evidence for this can be seen in the duality of $h \leftrightarrow J$
that that model possesses. This duality implies that the spectrum is
identical for $h \leftrightarrow J$, and as the strong transverse
field limit has a well defined gap, the ferromagnetic limit does too.

%%%%%%%%%%%%%%%%%%%%%%%%%%%%%%%%%%%%%%%%%%%%%%%%%%%%%%%%%%%%%%%%%%%%%%%%%%%%%
\section{Numerical results}
\label{sec:QMCres}

%Phase transition: Talk about and present data for 3-reg-3-XOR 
%first order (will have to present Binder here), move to talk about
%how this motivates perhaps other models on this lattice. Move to 2-reg-2
%and 1-c, talk about pattern of circle with offshoots, support with numeric

In this Section, we apply the QAA protocol to the lattice models for
$k$-regular $k$-XORSAT we introduced above. We simulate QAA via
projector Quantum Monte Carlo (QMC) simulations to investigate the
transitions encountered upon varying the transverse field. Our
calculations are set up following the style of simulations for TFI
models~\cite{QMC1,QMC2} due to the similarity in the Hamiltonians and
they are able to access ground state expectation values for various
observables. It is known that for glassy systems or at first-order
transitions, this style of QMC suffers from long equilibration times
and non-ergodic behavior~\cite{QMCF1}. Cluster updates using larger
objects made out of multiple spins, which have been found to be useful
for particular frustrated Ising antiferromagnets~\cite{BD}, were found
to be rejection-prone for our 3-XORSAT model due to the three-body
terms that make up CNOTs. This limits the sizes of the lattices we can
simulate and also how deep our simulations can reach into the
ferromagnetic phase. We use a variety of local and replica exchange
updates~\cite{ParrT1,ParrT2} to speed up the algorithm. Details of the
implementation of the replica exchange method can be found in
Ref.~\onlinecite{ParrT2}, where it has been applied to a similar model.
We will study our models in the context of continuous and first-order
phase transitions, which show a diverging and a finite correlation length,
respectively. The energy gap is expected to close for both types of
transitions, but the scaling of the minimum gap with system size
differs, closing as $\sim N^{-\gamma}$ for the former and
$\sim e^{-N^\gamma}$ for the latter $(\gamma>0)$.

The lattice models of $k$-regular $k$-XORSAT with a transverse field,
defined by Eqs.~\eqref{eq:Hanneal} and~\eqref{eq:Hxorlatt}, show two
phases: a disordered phase in the limit of strong transverse field and
an ordered phase in the ferromagnetic (or classical) limit. To
simplify our analysis, we set $h=1$ and vary $J$ through the
transition between the two phases. To find the critical value of the
ferromagnetic coupling $J_c$ and the nature of the phase transition,
we use the Binder cumulant~\cite{BC1}
\be
U_m = \frac{3}{2} \bigg(1-\frac{1}{3} \frac{\langle M_z^4\rangle}
{\langle M_z^2\rangle^2}\bigg),
\ee
where $M_z$ is the magnetization of the system in the $z$ direction.
In the strong transverse field limit, the system is magnetized along
the $x$ axis, which implies that $M_z=0$, whereas in the ferromagnetic
limit all spins are aligned, leading to a saturation of magnetization.
$U_m$ is defined such that it evaluates to zero in the disordered
phase and unity in the ordered one. At the phase transition, $U_m$
goes from 0 to 1 within a window of $J$ that gets narrower as system
size increases. For a continuous transition, the behavior is
monotonically increasing in most cases, although exceptions are
known~\cite{BCE}, and for a first-order transition $U_m$ has a
negative peak at the critical point which diverges with system size as
$L^d$, where $d$ is the dimensionality of the system~\cite{BC2}.

The Binder cumulant is chosen here to be 
sensitive to $Z_2$ symmetry breaking,
which would be the symmetry usually studied in Ising systems. For the $k=3$
case, the boundary conditions enforce a single solution, meaning that the
system is not doubly degenerate and in this case, $U_m$ shows a non-zero
value on the disordered side before the transition (seen in Fig.~\ref{Fig3X}),
as the histogram favors net negative magnetization. This anomaly is 
also seen in the transverse field Ising chain with fixed boundary conditions 
similar to the ones studied.

%%%%%%%%%%%%%%%%%%%%%%%%%%%%%%%%%%%%%%%%%%%%%%%%%%%%%%%%%%%%%%%%%%%%%%%%%%%%%%
\begin{figure}[t]
\includegraphics[width=0.4\hsize]{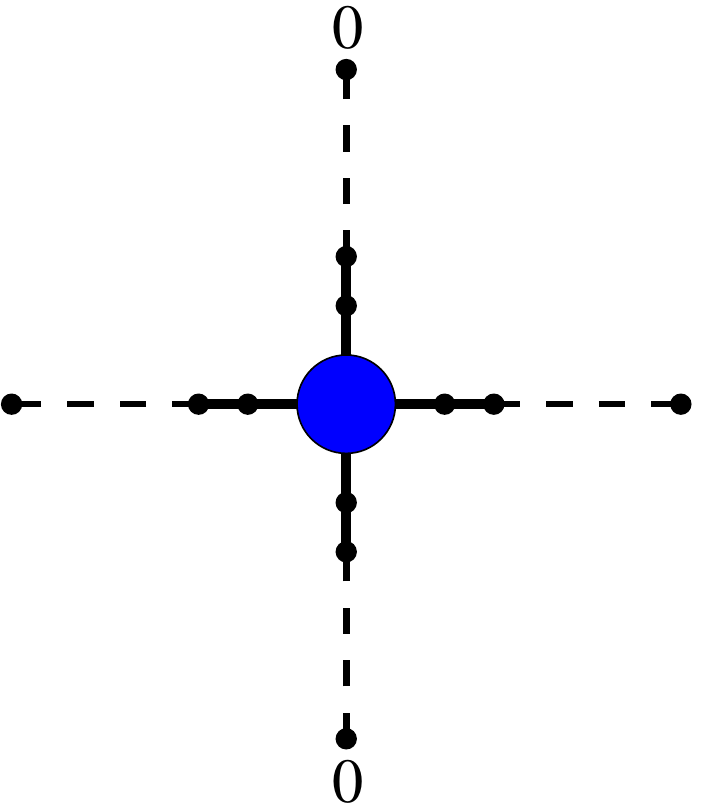}
\caption{Illustration of two TFI chains coupled at their center by a CNOT gate.
The boundary spins of the vertical chain are fixed.}
\label{fig:cross}
\end{figure}
%%%%%%%%%%%%%%%%%%%%%%%%%%%%%%%%%%%%%%%%%%%%%%%%%%%%%%%%%%%%%%%%%%%%%%%%%%%%%%

\begin{figure}[t]
\includegraphics[width=\hsize]{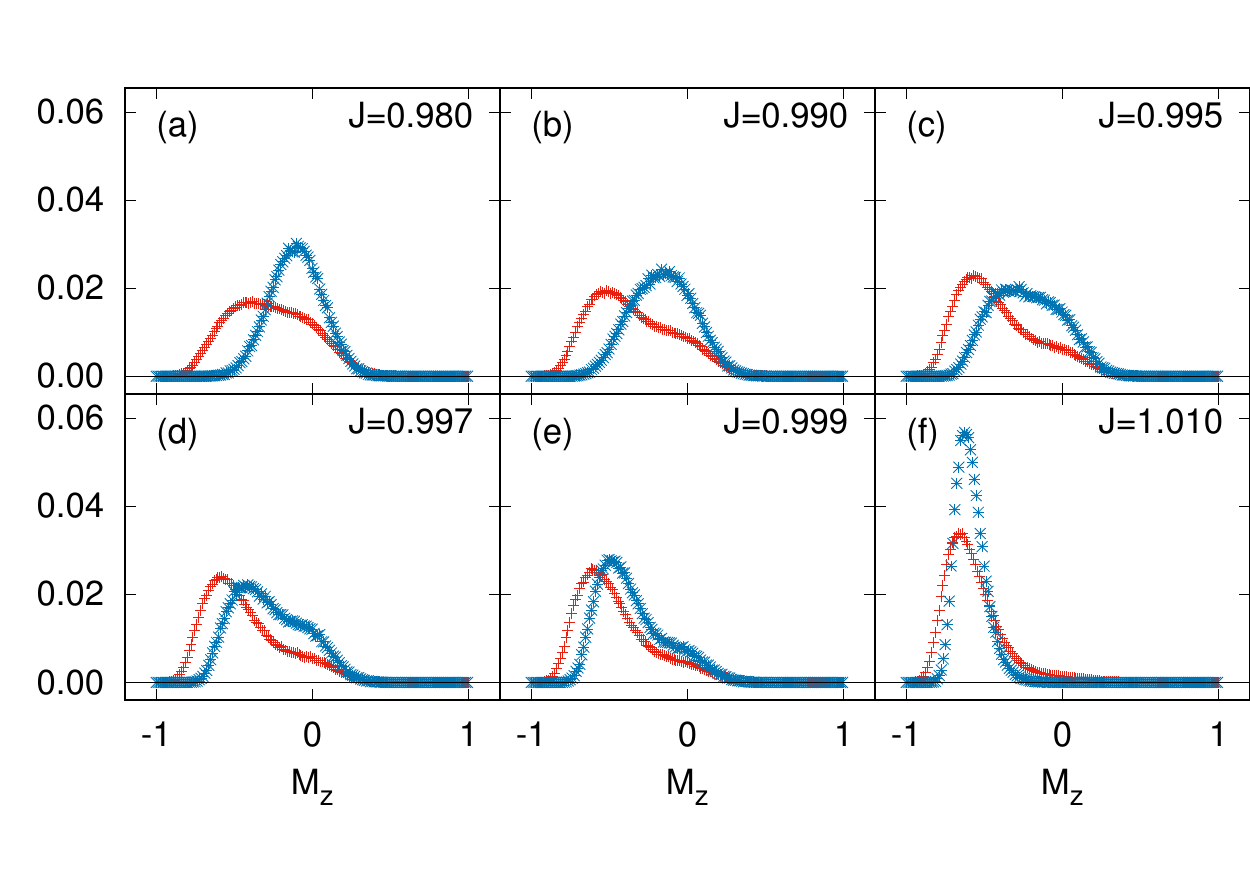}
\caption{Magnetization histograms for the transverse field system
  shown in Fig.~\ref{fig:cross} with 81 spins per chain (crosses) and 321 spins
  per chain (stars), as obtained
  from QMC calculations at $h=1$. As $J$ increases (a-f), the average
  magnetization switches values without developing a strong double peak 
  structure even for large size.}
\label{fig:C}
\end{figure}

We begin our study of the effectiveness of QAA on 2-regular 2-XORSAT
by first applying the algorithm to the building block of this system:
a single CNOT gate coupling long spin chains, as shown in
Fig.~\ref{fig:cross}. The CNOT coupling is expected to result in no
drastic change to the continuous phase transition of the two chains,
as it is a single defect in a large system. Numerical evidence for
this is shown in Fig.~\ref{fig:C}. Even for large systems of two
chains intersecting, the magnetization histogram shows no
coexistence of phases, indicative of a second-order transition. For
first-order transitions, on the other hand, the same histogram would
show a bimodal distribution, indicative of phase
coexistence~\cite{BC2}. We expect the same conclusion to extend to the
case of a large number of these ``cross''-linked chains connected
horizontally when the separation between CNOT gates is large, which is
the case for $k=2$, as shown in Fig.~\ref{fig:dw}. It is seen that
CNOTs interact only along a linear chain and correlation length growth
is controlled by the physics of the TFI chains connecting them. We
therefore expect a continuous phase transition at $h\sim J$.

%%%%%%%%%%%%%%%%%%%%%%%%%%%%%%%%%%%%%%%%%%%%%%%%%%%%%%%%%%%%%%%%%%%%%%%%%%%%%%
%%%%%%%%%%%%%%%%%%%%%%%%%%%%%%%%%%%%%%%%%%%%%%%%%%%%%%%%%%%%%%%%%%%%%%%%%%%%%%

%We first use this machinery to study the 2-regular2-XORSAT samples
%with two solutions (two is the minimum number of solutions as this
%problem is spin inversion symmetric), which places a global
%constraint on the locations of the CNOT gates. The random graph
%version of the 2-XORSAT with only two solutions can be
%written as a periodic Ising chain as we have mentioned
%earlier in this section with a transverse field, which would show a
%continuous transition. The two solutions that our realizations of
%$k=2$ lattices host are the state with all bit boundary spins pointing
%down and the one with all of them pointing up. It is worthwhile to
%note that the classical lattice ground states that these two solutions
%correspond to are themselves not spin symmetric due to the presence of
%CNOT gates in the Hamiltonian. Both lattice ground states have uniform
%magnetization (either up or down) for the spins along the bit
%(horizontal) lines and one has all spins down for the clause lines and
%the other has a particular pattern of spins in the clause lines
%specific to the realization.

\begin{figure}[t]
\includegraphics[width=\hsize]{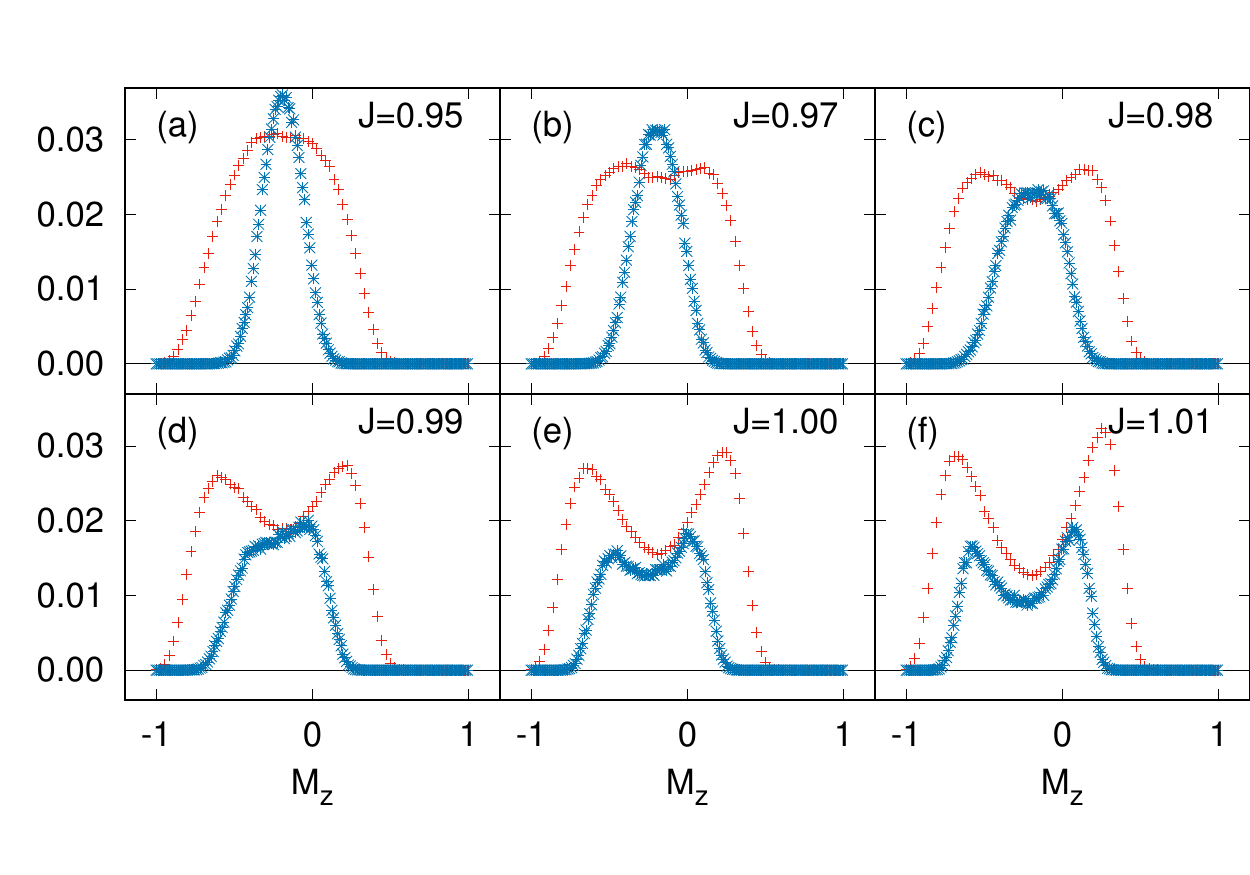}
\includegraphics[width=\hsize]{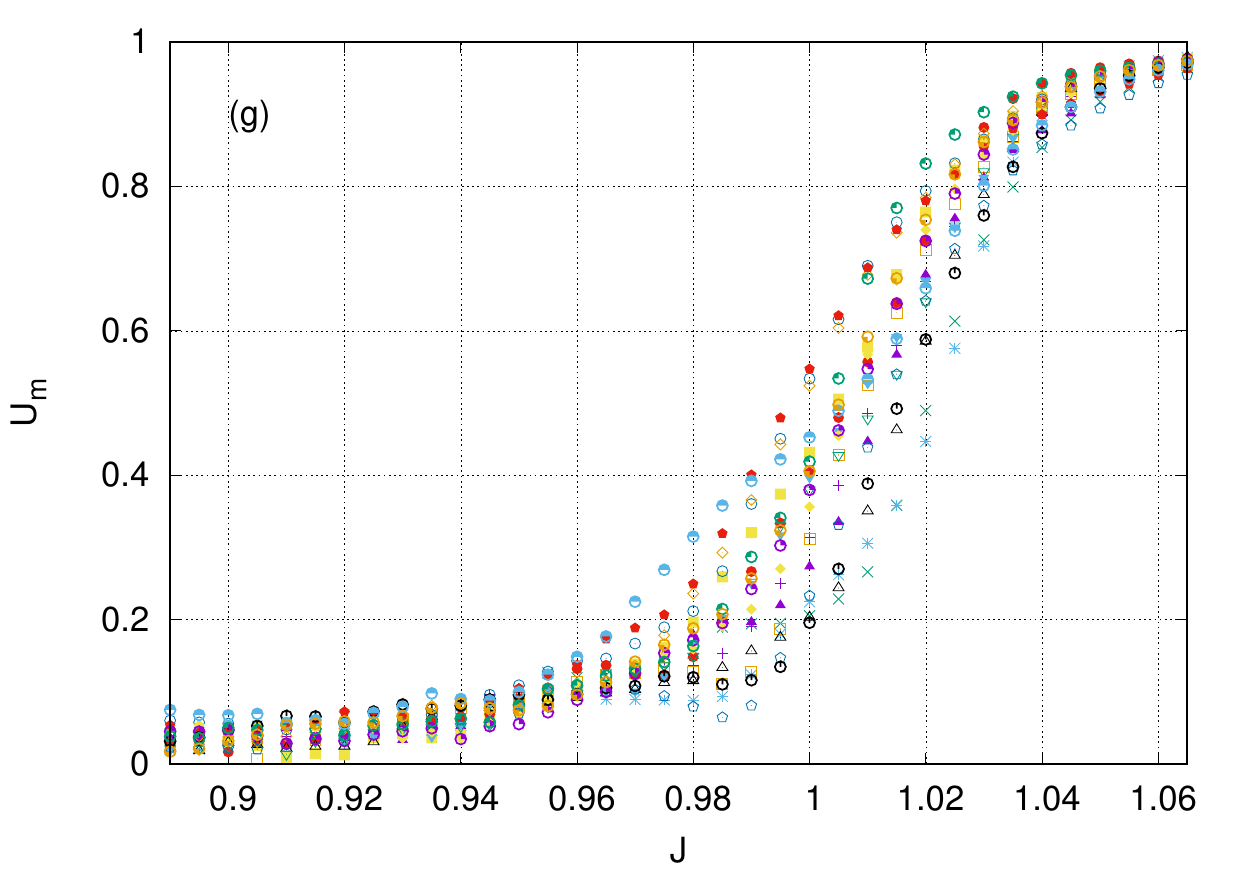}
\caption{(a-f) Magnetization histograms for a single realization of
  the 2-XORSAT lattice with size $6\times 6$ (crosses) and another
  with size $16\times 16$ (stars) for increasing $J$. The slight
  asymmetry in the two peaks of the histograms may be caused by
  non-ergodicity issues in the QMC simulation. (g) Binder cumulant for
  20 realizations of size $16\times 16$. In all panels $h=1$.}
\label{fig:2X}
\end{figure}

\begin{figure}[t]
\includegraphics[width=\hsize]{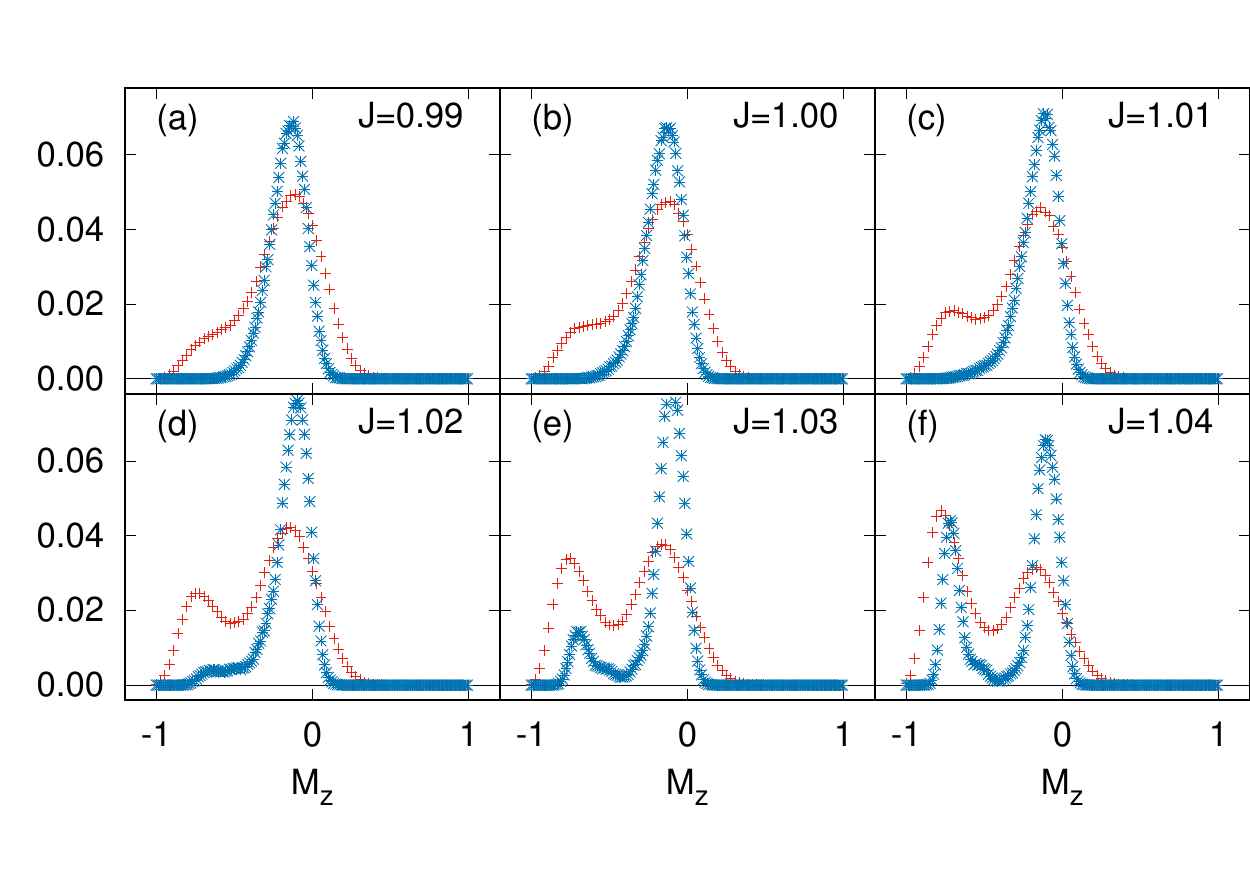}
\includegraphics[width=\hsize]{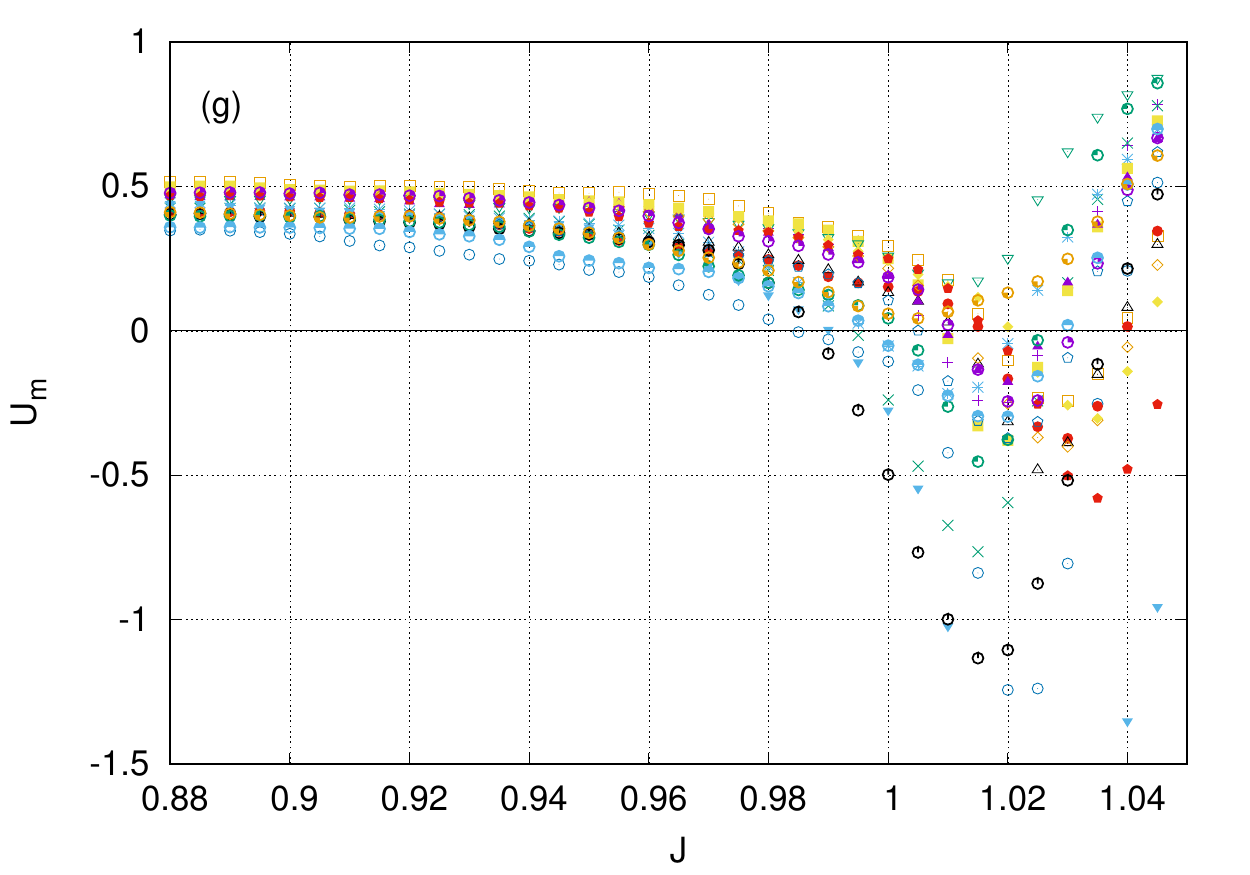}
\caption{(a-f) Magnetization histograms for a single realization of
  the 3-regular 3-XORSAT lattice with size $6\times 6$ (crosses) and
  another with size $16\times 16$ (stars) for increasing $J$. 
  %The slight asymmetry in the two peaks of the histograms is due to
  %non-ergodicity issues in the QMC simulation. 
  (g) Binder cumulant for
  20 realizations of size $16\times 16$. In all panels $h=1$.}
\label{Fig3X}
\end{figure}

We analyze 2-XORSAT instances by studying the behavior of $U_m$ and
the evolution of the magnetization histograms -- see
Fig.~\ref{fig:2X}. 20 realizations are used for each size and the
critical value of $J$ is found to be close to unity, which matches the
decoupled TFI chain value. 
For $k=2$, the two ground states correspond to
the configuration with all spins pointing down and the configuration with
all bit line spins pointing up and a complicated ordering of the clause
spins which is consistent with the solution and realization dependent.
Due to the non-trivial structure of the
ground states, $U_m$ has to be defined using only the variable spins,
as they form a spin-symmetric subset. This implies that the order
parameter we use for $U_m$ is the average $z$-magnetization squared of
only the row spins $i_1$ and $o_1$ at each $(x,y)$ position, which
make up half of the total spins. For all the realizations of all
lattice sizes, we found that $U_m$ never becomes negative and the
magnetization histograms show no sign of phase coexistence, indicating
a continuous transition, as seen in Fig.~\ref{fig:2X}.

%%%%%%%%%%%%%%%%%%%%%%%%%%%%%%%%%%%%%%%%%%%%%%%%%%%%%%%%%%%%%%%%%%%%%%%%%%%%%%
%%%%%%%%%%%%%%%%%%%%%%%%%%%%%%%%%%%%%%%%%%%%%%%%%%%%%%%%%%%%%%%%%%%%%%%%%%%%%%

The QAA protocol for 3-regular 3-XORSAT proceeds in the same vein as
for $k=2$. Using QMC simulations, we study the Binder cumulant using
the full magnetization (as $k=3$ has only one ground state) as a
function of ferromagnetic coupling $J$ for lattice sizes ranging from
$6\times 6$ to $16\times 16$. For each lattice size, we study 20
realizations of the 3-regular 3-XORSAT on the lattice with unique
solutions. We find that for most of the realizations and for all sizes
the Binder cumulant shows a negative peak. This can be seen in
Fig.~\ref{Fig3X}(g) for all realizations of a $16\times
16$ lattice. Due to lack of ergodicity typically seen at first order
phase transitions, we had to reject a large number of QMC simulations,
retaining only those where both phases are represented as some
simulations are not able to break out of the paramagnetic phase. This
in itself only indicates that we may have a first-order transition but
does not say so definitively. Showing that the negative peak in the
Binder cumulant scales as number of sites is taken to be definite
proof that the system is undergoing a first-order transition but we
are unable to perform this analysis due to insufficient data quality
and range of sizes. We omit error bars in Figs.~\ref{fig:2X}(g) and 
~\ref{Fig3X}(g) for clarity and as the error bars are small compared to
the spread of the Binder cumulant values for different realizations.

Fig.~\ref{Fig3X} shows the evolution of the histogram as a function
of $J$ for the 3-regular 3-XORSAT including the value of $J$ at which
$U_m$ is found to be minimum. The histogram indicates that there is a
coexistence of phases which sharpens with increasing size and the
phase transition can potentially develop into a first-order transition
in the thermodynamic limit. Our system sizes and data quality limit
our analysis, but for the system sizes we can access the first order
nature appears to persist to the largest system size
(Fig.~\ref{Fig3X}). It is important to stress here that although the
density of gates is vanishing in the thermodynamic limit, they are
placed in a correlated manner, and can hence define the criticality of
this model.

%%%%%%%%%%%%%%%%%%%%%%%%%%%%%%%%%%%%%%%%%%%%%%%%%%%%%%%%%%%%%%%%%%%%%%%%%%%%%%
%%%%%%%%%%%%%%%%%%%%%%%%%%%%%%%%%%%%%%%%%%%%%%%%%%%%%%%%%%%%%%%%%%%%%%%%%%%%%%

%%%%%%%%%%%%%%%%%%%%%%%%%%%%%%%%%%%%%%%%%%%%%%%%%%%%%%%%%%%%%%%%%%%%%%%%%%%%%
\section{Summary and conclusion}\label{Conc}

%Conclusions: Model has been used to show that PT is not always a problem
%and that there are interesting connections between complexity and
%physics. Maybe we can think of more examples like this ?

We introduced a statistical mechanics representation of the XORSAT
problem that recasts each instance as a planar grid of reversible
gates acting on bits that represent the Boolean variables of the
problem. The reason we chose this particular embedding of XORSAT is
that it lacks a classical thermodynamic phase transition. We studied
this system with quantum annealing and showed that it does not suffer
from the perturbation theory collapse found by Altshuler \emph{et
  al}.~\cite{Alt} at small transverse field strengths. We found that
3-regular 3-XORSAT displays a first-order transition at intermediate
values of the transverse field, implying that quantum annealing leads
to a time-to-solution that scales exponentially with the size of the
system. Our results also suggest that the physics of the phase
transition is determined not only by the density of CNOT defects, but
also by their correlations. Taken together, these results on this
alternative embedding of XORSAT reinforce the conclusion that both
thermal and quantum annealing, which are intrinsically local
approaches, can be inefficient in solving even simple problems (in
computational complexity class P).

%%%%%%%%%%%%%%%%%%%%%%%%%%%%%%%%%%%%%%%%%%%%%%%%%%%%%%%%%%%%%%%%%%%%%%%%%%%%%
\begin{acknowledgments}

We would like to thank Anders W. Sandvik and Jun Takahashi for
suggesting improvements to the algorithms and sharing valuable
technical expertise and Lei Zhang for valuable results from simulated
annealing. We would also like to thank Chris Baldwin, Phillip
Weinberg, and Itay Hen for useful discussions. The computational work
was performed using the Shared Computing Cluster administered by
Boston University's Research Computing Services. We used QuSpin for
checking the Quantum Monte Carlo simulation against exact
diagonalization calculations~\cite{Phil}. This work was partially
supported by the U.S. Department of Energy (DOE), Division of
Condensed Matter Physics and Materials Science, under Contract No.
DE-SC0019275 (PP, CC, and ERM).

%%%%%%%%%%%%%%%%%%%%%%%%%%%%%%%%%%%%%%%%%%%%%%%%%%%%%%%%%%%%%%%%%%%%%%%%%%%%%
\end{acknowledgments}
  
\appendix
\section{Mapping to the vertex model}
\label{sec:app}
%Stuff about mapping to prev paper with the diamond arrangement

\begin{figure}[ht]
\includegraphics[width=0.5\hsize]{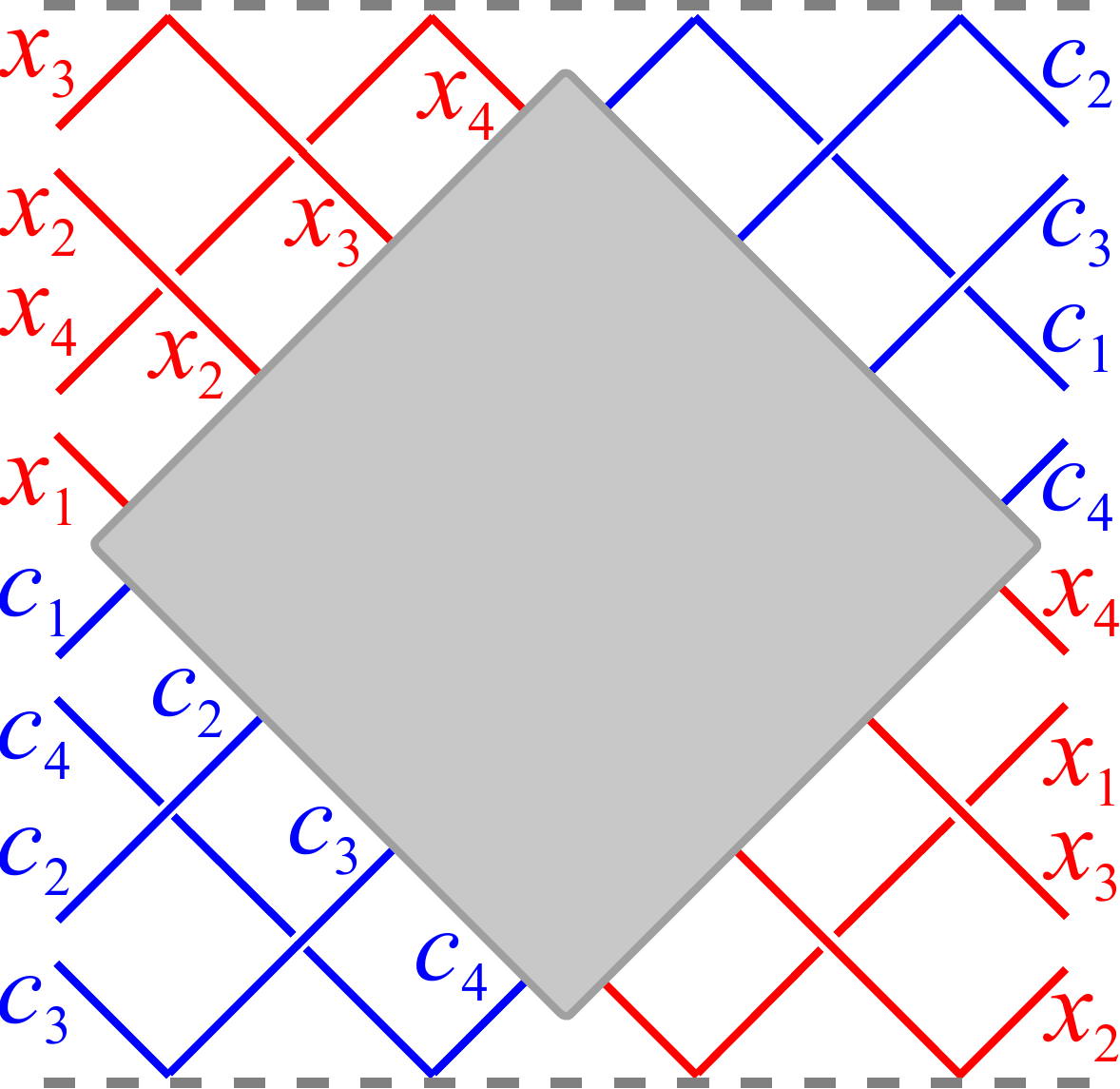}
\caption{Illustration of the mapping of the lattice formulation of
  $k$-XORSAT onto the general vertex model framework for computational
  problems introduced in Ref.~\onlinecite{Chamon2016}, here shown for a
  $4\times4$ lattice. The grey shaded area is the placeholder in which
  a lattice like the one shown in Fig.~\ref{fig:lat}(a) plugs into after
  a 45\textdegree rotation. In this embedding, clause bits (blue) are
  fixed and variable bits (red) are left free at both left and right
  boundaries, whereas all bit lines are ``reflected'' at the top and
  bottom boundaries (dashed lines).}
\label{fig:cir}
\end{figure}

Here we comment on the limitation of using thermal
annealing to reach the ground state of 3-regular 3-XORSAT in spite
of the absence of a bulk thermodynamic phase transition
~\cite{Chamon2016}. The simplest way to understand the slow thermal
relaxation into the ground state of the XORSAT model is to use the
recipe of Ref.~\onlinecite{Chamon2016} to embed XORSAT into the alternative
spin model shown in Fig.~\ref{fig:cir}. In this model, only half the
boundary spins are fixed in each
boundary, corresponding to the clause spins being fixed by a strong
field in Eq.~\eqref{eq:H2}, with the free bits (spins) on the rest of
the boundaries. Reaching the ground state of this model through
thermal annealing requires that information propagates between the
boundaries until all free spins on both boundaries are fixed. This
``mixed boundary condition'' case in which only partial initial
information is available on the input/output boundaries was already
studied in Ref.~\onlinecite{Chamon2016}. There it was found that thermal
annealing is ineffective in reconciling the non-local information
between the two boundaries, explaining the slow relaxation of the
lattice embedding of 3-regular 3-XORSAT.

%%%%%%%%%%%%%%%%%%%%%%%%%%%%%%%%%%%%%%%%%%%%%%%%%%%%%%%%%%%%%%%%%%%%%%%%%%%%%%
%%%%%%%%%%%%%%%%%%%%%%%%%%%%%%%%%%%%%%%%%%%%%%%%%%%%%%%%%%%%%%%%%%%%%%%%%%%%%%

%%%%%%%%%%%%%%%%%%%%%%%%%%%%%%%%%%%%%%%%%%%%%%%%%%%%%%%%%%%%%%%%%%%%%%%%%%%%%
\bibliography{citations}

%merlin.mbs apsrev4-1.bst 2010-07-25 4.21a (PWD, AO, DPC) hacked
%Control: key (0)
%Control: author (8) initials jnrlst
%Control: editor formatted (1) identically to author
%Control: production of article title (-1) disabled
%Control: page (0) single
%Control: year (1) truncated
%Control: production of eprint (0) enabled
\begin{thebibliography}{39}%
\makeatletter
\providecommand \@ifxundefined [1]{%
 \@ifx{#1\undefined}
}%
\providecommand \@ifnum [1]{%
 \ifnum #1\expandafter \@firstoftwo
 \else \expandafter \@secondoftwo
 \fi
}%
\providecommand \@ifx [1]{%
 \ifx #1\expandafter \@firstoftwo
 \else \expandafter \@secondoftwo
 \fi
}%
\providecommand \natexlab [1]{#1}%
\providecommand \enquote  [1]{``#1''}%
\providecommand \bibnamefont  [1]{#1}%
\providecommand \bibfnamefont [1]{#1}%
\providecommand \citenamefont [1]{#1}%
\providecommand \href@noop [0]{\@secondoftwo}%
\providecommand \href [0]{\begingroup \@sanitize@url \@href}%
\providecommand \@href[1]{\@@startlink{#1}\@@href}%
\providecommand \@@href[1]{\endgroup#1\@@endlink}%
\providecommand \@sanitize@url [0]{\catcode `\\12\catcode `\$12\catcode
  `\&12\catcode `\#12\catcode `\^12\catcode `\_12\catcode `\%12\relax}%
\providecommand \@@startlink[1]{}%
\providecommand \@@endlink[0]{}%
\providecommand \url  [0]{\begingroup\@sanitize@url \@url }%
\providecommand \@url [1]{\endgroup\@href {#1}{\urlprefix }}%
\providecommand \urlprefix  [0]{URL }%
\providecommand \Eprint [0]{\href }%
\providecommand \doibase [0]{http://dx.doi.org/}%
\providecommand \selectlanguage [0]{\@gobble}%
\providecommand \bibinfo  [0]{\@secondoftwo}%
\providecommand \bibfield  [0]{\@secondoftwo}%
\providecommand \translation [1]{[#1]}%
\providecommand \BibitemOpen [0]{}%
\providecommand \bibitemStop [0]{}%
\providecommand \bibitemNoStop [0]{.\EOS\space}%
\providecommand \EOS [0]{\spacefactor3000\relax}%
\providecommand \BibitemShut  [1]{\csname bibitem#1\endcsname}%
\let\auto@bib@innerbib\@empty
%</preamble>
\bibitem [{\citenamefont {Kirkpatrick}\ \emph {et~al.}(1983)\citenamefont
  {Kirkpatrick}, \citenamefont {Gellat~Jr.},\ and\ \citenamefont
  {Vecchi}}]{Kirkpatrick1983}%
  \BibitemOpen
  \bibfield  {author} {\bibinfo {author} {\bibfnamefont {S.}~\bibnamefont
  {Kirkpatrick}}, \bibinfo {author} {\bibfnamefont {C.~D.}\ \bibnamefont
  {Gellat~Jr.}}, \ and\ \bibinfo {author} {\bibfnamefont {M.~P.}\ \bibnamefont
  {Vecchi}},\ }\href {\doibase 10.1126/science.220.4598.671} {\bibfield
  {journal} {\bibinfo  {journal} {Science}\ }\textbf {\bibinfo {volume}
  {220}},\ \bibinfo {pages} {671} (\bibinfo {year} {1983})}\BibitemShut
  {NoStop}%
\bibitem [{\citenamefont {M{\'e}zard}\ \emph {et~al.}(2002)\citenamefont
  {M{\'e}zard}, \citenamefont {Parisi},\ and\ \citenamefont
  {Zecchina}}]{Zechina-Parisi-Mezard}%
  \BibitemOpen
  \bibfield  {author} {\bibinfo {author} {\bibfnamefont {M.}~\bibnamefont
  {M{\'e}zard}}, \bibinfo {author} {\bibfnamefont {G.}~\bibnamefont {Parisi}},
  \ and\ \bibinfo {author} {\bibfnamefont {R.}~\bibnamefont {Zecchina}},\
  }\href@noop {} {\bibfield  {journal} {\bibinfo  {journal} {Science}\ }\textbf
  {\bibinfo {volume} {297}},\ \bibinfo {pages} {812} (\bibinfo {year}
  {2002})}\BibitemShut {NoStop}%
\bibitem [{\citenamefont {M{\'e}zard}\ and\ \citenamefont
  {Montanari}(2009)}]{Montanari-and-Mezard-book}%
  \BibitemOpen
  \bibfield  {author} {\bibinfo {author} {\bibfnamefont {M.}~\bibnamefont
  {M{\'e}zard}}\ and\ \bibinfo {author} {\bibfnamefont {A.}~\bibnamefont
  {Montanari}},\ }\href@noop {} {\emph {\bibinfo {title} {Information, physics,
  and computation}}}\ (\bibinfo  {publisher} {Oxford University Press},\
  \bibinfo {year} {2009})\BibitemShut {NoStop}%
\bibitem [{\citenamefont {Ricci-Tersenghi}(2010)}]{Ricci-Tersenghi2010}%
  \BibitemOpen
  \bibfield  {author} {\bibinfo {author} {\bibfnamefont {F.}~\bibnamefont
  {Ricci-Tersenghi}},\ }\href {\doibase 10.1126/science.1189804} {\bibfield
  {journal} {\bibinfo  {journal} {Science}\ }\textbf {\bibinfo {volume}
  {330}},\ \bibinfo {pages} {1639} (\bibinfo {year} {2010})}\BibitemShut
  {NoStop}%
\bibitem [{\citenamefont {Chamon}\ \emph {et~al.}(2017)\citenamefont {Chamon},
  \citenamefont {Mucciolo}, \citenamefont {Ruckenstein},\ and\ \citenamefont
  {Yang}}]{Chamon2016}%
  \BibitemOpen
  \bibfield  {author} {\bibinfo {author} {\bibfnamefont {C.}~\bibnamefont
  {Chamon}}, \bibinfo {author} {\bibfnamefont {E.~R.}\ \bibnamefont
  {Mucciolo}}, \bibinfo {author} {\bibfnamefont {A.~E.}\ \bibnamefont
  {Ruckenstein}}, \ and\ \bibinfo {author} {\bibfnamefont {Z.-C.}\ \bibnamefont
  {Yang}},\ }\href {\doibase 10.1038/ncomms15303} {\bibfield  {journal}
  {\bibinfo  {journal} {Nat. Commun.}\ }\textbf {\bibinfo {volume} {8}},\
  \bibinfo {pages} {15303} (\bibinfo {year} {2017})}\BibitemShut {NoStop}%
\bibitem [{\citenamefont {Zhang}\ \emph {et~al.}(2018)\citenamefont {Zhang},
  \citenamefont {Kourtis}, \citenamefont {Chamon}, \citenamefont {Mucciolo},\
  and\ \citenamefont {Ruckenstein}}]{Lei}%
  \BibitemOpen
  \bibfield  {author} {\bibinfo {author} {\bibfnamefont {L.}~\bibnamefont
  {Zhang}}, \bibinfo {author} {\bibfnamefont {S.}~\bibnamefont {Kourtis}},
  \bibinfo {author} {\bibfnamefont {C.}~\bibnamefont {Chamon}}, \bibinfo
  {author} {\bibfnamefont {E.~R.}\ \bibnamefont {Mucciolo}}, \ and\ \bibinfo
  {author} {\bibfnamefont {A.~E.}\ \bibnamefont {Ruckenstein}},\ }\href@noop {}
  {\bibfield  {journal} {\bibinfo  {journal} {preprint arXiv:1812.01621}\ }
  (\bibinfo {year} {2018})}\BibitemShut {NoStop}%
\bibitem [{\citenamefont {Yang}(2013)}]{GE}%
  \BibitemOpen
  \bibfield  {author} {\bibinfo {author} {\bibfnamefont {Y.}~\bibnamefont
  {Yang}},\ }\href@noop {} {\bibfield  {journal} {\bibinfo  {journal} {Math.
  Comput. Educ.}\ }\textbf {\bibinfo {volume} {47}},\ \bibinfo {pages} {224}
  (\bibinfo {year} {2013})}\BibitemShut {NoStop}%
\bibitem [{\citenamefont {Boixo}(2014)}]{TNat}%
  \BibitemOpen
  \bibfield  {author} {\bibinfo {author} {\bibfnamefont {S.}~\bibnamefont
  {Boixo}},\ }\href@noop {} {\bibfield  {journal} {\bibinfo  {journal} {Nat.
  Phys.}\ }\textbf {\bibinfo {volume} {10}},\ \bibinfo {pages} {218} (\bibinfo
  {year} {2014})}\BibitemShut {NoStop}%
\bibitem [{\citenamefont {Farhi}\ \emph {et~al.}(2000)\citenamefont {Farhi},
  \citenamefont {Goldstone}, \citenamefont {Gutmann},\ and\ \citenamefont
  {Sipser}}]{EFold}%
  \BibitemOpen
  \bibfield  {author} {\bibinfo {author} {\bibfnamefont {E.}~\bibnamefont
  {Farhi}}, \bibinfo {author} {\bibfnamefont {J.}~\bibnamefont {Goldstone}},
  \bibinfo {author} {\bibfnamefont {S.}~\bibnamefont {Gutmann}}, \ and\
  \bibinfo {author} {\bibfnamefont {M.}~\bibnamefont {Sipser}},\ }\href@noop {}
  {\bibfield  {journal} {\bibinfo  {journal} {arXiv:quant-ph/0001106}\ }
  (\bibinfo {year} {2000})}\BibitemShut {NoStop}%
\bibitem [{\citenamefont {Farhi}\ \emph {et~al.}(2012)\citenamefont {Farhi},
  \citenamefont {Gosset}, \citenamefont {Hen}, \citenamefont {Sandvik},
  \citenamefont {Shor}, \citenamefont {Young},\ and\ \citenamefont
  {Zamponi}}]{EF}%
  \BibitemOpen
  \bibfield  {author} {\bibinfo {author} {\bibfnamefont {E.}~\bibnamefont
  {Farhi}}, \bibinfo {author} {\bibfnamefont {D.}~\bibnamefont {Gosset}},
  \bibinfo {author} {\bibfnamefont {I.}~\bibnamefont {Hen}}, \bibinfo {author}
  {\bibfnamefont {A.~W.}\ \bibnamefont {Sandvik}}, \bibinfo {author}
  {\bibfnamefont {P.}~\bibnamefont {Shor}}, \bibinfo {author} {\bibfnamefont
  {A.~P.}\ \bibnamefont {Young}}, \ and\ \bibinfo {author} {\bibfnamefont
  {F.}~\bibnamefont {Zamponi}},\ }\href@noop {} {\bibfield  {journal} {\bibinfo
   {journal} {Phys. Rev. A}\ }\textbf {\bibinfo {volume} {86}},\ \bibinfo
  {pages} {052334} (\bibinfo {year} {2012})}\BibitemShut {NoStop}%
\bibitem [{\citenamefont {Kadowaki}\ and\ \citenamefont
  {Nishimori}(1998)}]{QATIM}%
  \BibitemOpen
  \bibfield  {author} {\bibinfo {author} {\bibfnamefont {T.}~\bibnamefont
  {Kadowaki}}\ and\ \bibinfo {author} {\bibfnamefont {H.}~\bibnamefont
  {Nishimori}},\ }\href {\doibase 10.1103/PhysRevE.58.5355} {\bibfield
  {journal} {\bibinfo  {journal} {Phys. Rev. E}\ }\textbf {\bibinfo {volume}
  {58}},\ \bibinfo {pages} {5355} (\bibinfo {year} {1998})}\BibitemShut
  {NoStop}%
\bibitem [{\citenamefont {Albash}\ and\ \citenamefont
  {Lidar}(2018)}]{QAscaling}%
  \BibitemOpen
  \bibfield  {author} {\bibinfo {author} {\bibfnamefont {T.}~\bibnamefont
  {Albash}}\ and\ \bibinfo {author} {\bibfnamefont {D.~A.}\ \bibnamefont
  {Lidar}},\ }\href@noop {} {\bibfield  {journal} {\bibinfo  {journal}
  {Physical Review X}\ }\textbf {\bibinfo {volume} {8}},\ \bibinfo {pages}
  {031016} (\bibinfo {year} {2018})}\BibitemShut {NoStop}%
\bibitem [{\citenamefont {Farhi}\ \emph
  {et~al.}(2001{\natexlab{a}})\citenamefont {Farhi}, \citenamefont {Goldstone},
  \citenamefont {Gutmann}, \citenamefont {Lapan}, \citenamefont {Lundgren},\
  and\ \citenamefont {Preda}}]{EFscience}%
  \BibitemOpen
  \bibfield  {author} {\bibinfo {author} {\bibfnamefont {E.}~\bibnamefont
  {Farhi}}, \bibinfo {author} {\bibfnamefont {J.}~\bibnamefont {Goldstone}},
  \bibinfo {author} {\bibfnamefont {S.}~\bibnamefont {Gutmann}}, \bibinfo
  {author} {\bibfnamefont {J.}~\bibnamefont {Lapan}}, \bibinfo {author}
  {\bibfnamefont {A.}~\bibnamefont {Lundgren}}, \ and\ \bibinfo {author}
  {\bibfnamefont {D.}~\bibnamefont {Preda}},\ }\href@noop {} {\bibfield
  {journal} {\bibinfo  {journal} {Science}\ }\textbf {\bibinfo {volume}
  {292}},\ \bibinfo {pages} {472} (\bibinfo {year}
  {2001}{\natexlab{a}})}\BibitemShut {NoStop}%
\bibitem [{\citenamefont {Altshuler}\ \emph {et~al.}(2010)\citenamefont
  {Altshuler}, \citenamefont {Krovi},\ and\ \citenamefont {Roland}}]{Alt}%
  \BibitemOpen
  \bibfield  {author} {\bibinfo {author} {\bibfnamefont {B.}~\bibnamefont
  {Altshuler}}, \bibinfo {author} {\bibfnamefont {H.}~\bibnamefont {Krovi}}, \
  and\ \bibinfo {author} {\bibfnamefont {J.}~\bibnamefont {Roland}},\
  }\href@noop {} {\bibfield  {journal} {\bibinfo  {journal} {Proc. Natl. Acad.
  Sci. USA}\ }\textbf {\bibinfo {volume} {107}},\ \bibinfo {pages} {12446}
  (\bibinfo {year} {2010})}\BibitemShut {NoStop}%
\bibitem [{\citenamefont {Farhi}\ \emph
  {et~al.}(2001{\natexlab{b}})\citenamefont {Farhi}, \citenamefont {Goldstone},
  \citenamefont {Gutmann}, \citenamefont {Lapan}, \citenamefont {Lundgren},\
  and\ \citenamefont {Preda}}]{Farhi2001}%
  \BibitemOpen
  \bibfield  {author} {\bibinfo {author} {\bibfnamefont {E.}~\bibnamefont
  {Farhi}}, \bibinfo {author} {\bibfnamefont {J.}~\bibnamefont {Goldstone}},
  \bibinfo {author} {\bibfnamefont {S.}~\bibnamefont {Gutmann}}, \bibinfo
  {author} {\bibfnamefont {J.}~\bibnamefont {Lapan}}, \bibinfo {author}
  {\bibfnamefont {A.}~\bibnamefont {Lundgren}}, \ and\ \bibinfo {author}
  {\bibfnamefont {D.}~\bibnamefont {Preda}},\ }\href {\doibase
  10.1126/science.1057726} {\bibfield  {journal} {\bibinfo  {journal}
  {Science}\ }\textbf {\bibinfo {volume} {292}},\ \bibinfo {pages} {472}
  (\bibinfo {year} {2001}{\natexlab{b}})}\BibitemShut {NoStop}%
\bibitem [{\citenamefont {Wannier}(1965)}]{Qad1}%
  \BibitemOpen
  \bibfield  {author} {\bibinfo {author} {\bibfnamefont {G.~H.}\ \bibnamefont
  {Wannier}},\ }\href@noop {} {\bibfield  {journal} {\bibinfo  {journal}
  {Physics Physique Fizika}\ }\textbf {\bibinfo {volume} {1}},\ \bibinfo
  {pages} {251} (\bibinfo {year} {1965})}\BibitemShut {NoStop}%
\bibitem [{\citenamefont {Kato}(1950)}]{Qad2}%
  \BibitemOpen
  \bibfield  {author} {\bibinfo {author} {\bibfnamefont {T.}~\bibnamefont
  {Kato}},\ }\href@noop {} {\bibfield  {journal} {\bibinfo  {journal} {J. Phys.
  Soc. Japan}\ }\textbf {\bibinfo {volume} {5}},\ \bibinfo {pages} {435}
  (\bibinfo {year} {1950})}\BibitemShut {NoStop}%
\bibitem [{\citenamefont {Sondhi}\ \emph {et~al.}(1997)\citenamefont {Sondhi},
  \citenamefont {Girvin}, \citenamefont {Carini},\ and\ \citenamefont
  {Shahar}}]{SQPT}%
  \BibitemOpen
  \bibfield  {author} {\bibinfo {author} {\bibfnamefont {S.~L.}\ \bibnamefont
  {Sondhi}}, \bibinfo {author} {\bibfnamefont {S.~M.}\ \bibnamefont {Girvin}},
  \bibinfo {author} {\bibfnamefont {J.~P.}\ \bibnamefont {Carini}}, \ and\
  \bibinfo {author} {\bibfnamefont {D.}~\bibnamefont {Shahar}},\ }\href@noop {}
  {\bibfield  {journal} {\bibinfo  {journal} {Rev. Mod. Phys.}\ }\textbf
  {\bibinfo {volume} {69}},\ \bibinfo {pages} {315} (\bibinfo {year}
  {1997})}\BibitemShut {NoStop}%
\bibitem [{\citenamefont {Laumann}\ \emph {et~al.}(2012)\citenamefont
  {Laumann}, \citenamefont {Moessner}, \citenamefont {Scardicchio},\ and\
  \citenamefont {Sondhi}}]{FQPTgap}%
  \BibitemOpen
  \bibfield  {author} {\bibinfo {author} {\bibfnamefont {C.~R.}\ \bibnamefont
  {Laumann}}, \bibinfo {author} {\bibfnamefont {R.}~\bibnamefont {Moessner}},
  \bibinfo {author} {\bibfnamefont {A.}~\bibnamefont {Scardicchio}}, \ and\
  \bibinfo {author} {\bibfnamefont {S.~L.}\ \bibnamefont {Sondhi}},\
  }\href@noop {} {\bibfield  {journal} {\bibinfo  {journal} {Phys. Rev. Lett.}\
  }\textbf {\bibinfo {volume} {109}},\ \bibinfo {pages} {030502} (\bibinfo
  {year} {2012})}\BibitemShut {NoStop}%
\bibitem [{\citenamefont {Ricci-Tersenghi}\ \emph {et~al.}(2001)\citenamefont
  {Ricci-Tersenghi}, \citenamefont {Weigt},\ and\ \citenamefont
  {Zecchina}}]{Ricci-Tersenghi2001}%
  \BibitemOpen
  \bibfield  {author} {\bibinfo {author} {\bibfnamefont {F.}~\bibnamefont
  {Ricci-Tersenghi}}, \bibinfo {author} {\bibfnamefont {M.}~\bibnamefont
  {Weigt}}, \ and\ \bibinfo {author} {\bibfnamefont {R.}~\bibnamefont
  {Zecchina}},\ }\href {\doibase 10.1103/PhysRevE.63.026702} {\bibfield
  {journal} {\bibinfo  {journal} {Phys. Rev. E}\ }\textbf {\bibinfo {volume}
  {63}},\ \bibinfo {pages} {026702} (\bibinfo {year} {2001})}\BibitemShut
  {NoStop}%
\bibitem [{\citenamefont {Johnson}\ and\ \citenamefont {Garey}(1979)}]{Xref}%
  \BibitemOpen
  \bibfield  {author} {\bibinfo {author} {\bibfnamefont {D.~S.}\ \bibnamefont
  {Johnson}}\ and\ \bibinfo {author} {\bibfnamefont {M.~R.}\ \bibnamefont
  {Garey}},\ }\href@noop {} {\ \textbf {\bibinfo {volume} {1}} (\bibinfo {year}
  {1979})}\BibitemShut {NoStop}%
\bibitem [{\citenamefont {Haanp{\"{a}}{\"{a}}}\ \emph
  {et~al.}(2006)\citenamefont {Haanp{\"{a}}{\"{a}}}, \citenamefont
  {J{\"{a}}rvisalo}, \citenamefont {Kaski},\ and\ \citenamefont
  {Niemel{\"{a}}}}]{Haanp2006}%
  \BibitemOpen
  \bibfield  {author} {\bibinfo {author} {\bibfnamefont {H.}~\bibnamefont
  {Haanp{\"{a}}{\"{a}}}}, \bibinfo {author} {\bibfnamefont {M.}~\bibnamefont
  {J{\"{a}}rvisalo}}, \bibinfo {author} {\bibfnamefont {P.}~\bibnamefont
  {Kaski}}, \ and\ \bibinfo {author} {\bibfnamefont {I.}~\bibnamefont
  {Niemel{\"{a}}}},\ }\href@noop {} {\bibfield  {journal} {\bibinfo  {journal}
  {Journal on Satisfiability, Boolean Modeling and Computation}\ }\textbf
  {\bibinfo {volume} {2}},\ \bibinfo {pages} {27} (\bibinfo {year}
  {2006})}\BibitemShut {NoStop}%
\bibitem [{\citenamefont {Jia}\ \emph {et~al.}(2005)\citenamefont {Jia},
  \citenamefont {Moore},\ and\ \citenamefont {Selman}}]{Jia2005}%
  \BibitemOpen
  \bibfield  {author} {\bibinfo {author} {\bibfnamefont {H.}~\bibnamefont
  {Jia}}, \bibinfo {author} {\bibfnamefont {C.}~\bibnamefont {Moore}}, \ and\
  \bibinfo {author} {\bibfnamefont {B.}~\bibnamefont {Selman}},\ }in\ \href
  {\doibase 10.1007/11527695_16} {\emph {\bibinfo {booktitle} {Theory and
  Applications of Satisfiability Testing, 7th International Conference, SAT
  2004}}}\ (\bibinfo {year} {2005})\ pp.\ \bibinfo {pages}
  {199--210}\BibitemShut {NoStop}%
\bibitem [{\citenamefont {Guidetti}\ and\ \citenamefont {Young}(2011)}]{XNL2}%
  \BibitemOpen
  \bibfield  {author} {\bibinfo {author} {\bibfnamefont {M.}~\bibnamefont
  {Guidetti}}\ and\ \bibinfo {author} {\bibfnamefont {A.~P.}\ \bibnamefont
  {Young}},\ }\href@noop {} {\bibfield  {journal} {\bibinfo  {journal} {Phys.
  Rev. E}\ }\textbf {\bibinfo {volume} {84}},\ \bibinfo {pages} {011102}
  (\bibinfo {year} {2011})}\BibitemShut {NoStop}%
\bibitem [{\citenamefont {J{\"{o}}rg}\ \emph {et~al.}(2010)\citenamefont
  {J{\"{o}}rg}, \citenamefont {Krzakala}, \citenamefont {Semerjian},\ and\
  \citenamefont {Zamponi}}]{Jorg2010}%
  \BibitemOpen
  \bibfield  {author} {\bibinfo {author} {\bibfnamefont {T.}~\bibnamefont
  {J{\"{o}}rg}}, \bibinfo {author} {\bibfnamefont {F.}~\bibnamefont
  {Krzakala}}, \bibinfo {author} {\bibfnamefont {G.}~\bibnamefont {Semerjian}},
  \ and\ \bibinfo {author} {\bibfnamefont {F.}~\bibnamefont {Zamponi}},\ }\href
  {\doibase 10.1103/PhysRevLett.104.207206} {\bibfield  {journal} {\bibinfo
  {journal} {Phys. Rev. Lett.}\ }\textbf {\bibinfo {volume} {104}},\ \bibinfo
  {pages} {207206} (\bibinfo {year} {2010})}\BibitemShut {NoStop}%
\bibitem [{\citenamefont {Pfeuty}(1970)}]{PTFI}%
  \BibitemOpen
  \bibfield  {author} {\bibinfo {author} {\bibfnamefont {P.}~\bibnamefont
  {Pfeuty}},\ }\href@noop {} {\bibfield  {journal} {\bibinfo  {journal} {Annals
  Phys.}\ }\textbf {\bibinfo {volume} {57}},\ \bibinfo {pages} {79} (\bibinfo
  {year} {1970})}\BibitemShut {NoStop}%
\bibitem [{\citenamefont {Flory}\ and\ \citenamefont {Rehner~Jr}(1943)}]{VR}%
  \BibitemOpen
  \bibfield  {author} {\bibinfo {author} {\bibfnamefont {P.~J.}\ \bibnamefont
  {Flory}}\ and\ \bibinfo {author} {\bibfnamefont {J.}~\bibnamefont
  {Rehner~Jr}},\ }\href@noop {} {\bibfield  {journal} {\bibinfo  {journal} {J.
  Chem. Phys.}\ }\textbf {\bibinfo {volume} {11}},\ \bibinfo {pages} {512}
  (\bibinfo {year} {1943})}\BibitemShut {NoStop}%
\bibitem [{\citenamefont {Knysh}\ and\ \citenamefont
  {Smelyanskiy}(2010)}]{Knysh2010}%
  \BibitemOpen
  \bibfield  {author} {\bibinfo {author} {\bibfnamefont {S.}~\bibnamefont
  {Knysh}}\ and\ \bibinfo {author} {\bibfnamefont {V.}~\bibnamefont
  {Smelyanskiy}},\ }\href {http://arxiv.org/abs/1005.3011} {\bibfield
  {journal} {\bibinfo  {journal} {arXiv:1005.3011}\ } (\bibinfo {year}
  {2010})},\ \Eprint {http://arxiv.org/abs/1005.3011} {arXiv:1005.3011}
  \BibitemShut {NoStop}%
\bibitem [{\citenamefont {Suzuki}\ \emph {et~al.}(2012)\citenamefont {Suzuki},
  \citenamefont {Inoue},\ and\ \citenamefont {Chakrabarti}}]{IMPT}%
  \BibitemOpen
  \bibfield  {author} {\bibinfo {author} {\bibfnamefont {S.}~\bibnamefont
  {Suzuki}}, \bibinfo {author} {\bibfnamefont {J.-i.}\ \bibnamefont {Inoue}}, \
  and\ \bibinfo {author} {\bibfnamefont {B.~K.}\ \bibnamefont {Chakrabarti}},\
  }\href@noop {} {\emph {\bibinfo {title} {Quantum Ising phases and transitions
  in transverse Ising models}}},\ Vol.\ \bibinfo {volume} {862}\ (\bibinfo
  {publisher} {Springer},\ \bibinfo {year} {2012})\BibitemShut {NoStop}%
\bibitem [{\citenamefont {Sandvik}(2003)}]{QMC1}%
  \BibitemOpen
  \bibfield  {author} {\bibinfo {author} {\bibfnamefont {A.~W.}\ \bibnamefont
  {Sandvik}},\ }\href@noop {} {\bibfield  {journal} {\bibinfo  {journal} {Phys.
  Rev. E}\ }\textbf {\bibinfo {volume} {68}},\ \bibinfo {pages} {056701}
  (\bibinfo {year} {2003})}\BibitemShut {NoStop}%
\bibitem [{\citenamefont {Liu}\ \emph {et~al.}(2013)\citenamefont {Liu},
  \citenamefont {Polkovnikov},\ and\ \citenamefont {Sandvik}}]{QMC2}%
  \BibitemOpen
  \bibfield  {author} {\bibinfo {author} {\bibfnamefont {C.-W.}\ \bibnamefont
  {Liu}}, \bibinfo {author} {\bibfnamefont {A.}~\bibnamefont {Polkovnikov}}, \
  and\ \bibinfo {author} {\bibfnamefont {A.~W.}\ \bibnamefont {Sandvik}},\
  }\href@noop {} {\bibfield  {journal} {\bibinfo  {journal} {Phys. Rev. B}\
  }\textbf {\bibinfo {volume} {87}},\ \bibinfo {pages} {174302} (\bibinfo
  {year} {2013})}\BibitemShut {NoStop}%
\bibitem [{\citenamefont {Brady}\ and\ \citenamefont {van Dam}(2016)}]{QMCF1}%
  \BibitemOpen
  \bibfield  {author} {\bibinfo {author} {\bibfnamefont {L.~T.}\ \bibnamefont
  {Brady}}\ and\ \bibinfo {author} {\bibfnamefont {W.}~\bibnamefont {van
  Dam}},\ }\href@noop {} {\bibfield  {journal} {\bibinfo  {journal} {Phys. Rev.
  A}\ }\textbf {\bibinfo {volume} {93}},\ \bibinfo {pages} {032304} (\bibinfo
  {year} {2016})}\BibitemShut {NoStop}%
\bibitem [{\citenamefont {Biswas}\ and\ \citenamefont {Damle}(2018)}]{BD}%
  \BibitemOpen
  \bibfield  {author} {\bibinfo {author} {\bibfnamefont {S.}~\bibnamefont
  {Biswas}}\ and\ \bibinfo {author} {\bibfnamefont {K.}~\bibnamefont {Damle}},\
  }\href@noop {} {\bibfield  {journal} {\bibinfo  {journal} {Phys. Rev. B}\
  }\textbf {\bibinfo {volume} {97}},\ \bibinfo {pages} {085114} (\bibinfo
  {year} {2018})}\BibitemShut {NoStop}%
\bibitem [{\citenamefont {Hukushima}\ and\ \citenamefont
  {Nemoto}(1996)}]{ParrT1}%
  \BibitemOpen
  \bibfield  {author} {\bibinfo {author} {\bibfnamefont {K.}~\bibnamefont
  {Hukushima}}\ and\ \bibinfo {author} {\bibfnamefont {K.}~\bibnamefont
  {Nemoto}},\ }\href@noop {} {\bibfield  {journal} {\bibinfo  {journal} {J.
  Phys. Soc. Japan}\ }\textbf {\bibinfo {volume} {65}},\ \bibinfo {pages}
  {1604} (\bibinfo {year} {1996})}\BibitemShut {NoStop}%
\bibitem [{\citenamefont {Takahashi}\ and\ \citenamefont
  {Hukushima}(2019)}]{ParrT2}%
  \BibitemOpen
  \bibfield  {author} {\bibinfo {author} {\bibfnamefont {J.}~\bibnamefont
  {Takahashi}}\ and\ \bibinfo {author} {\bibfnamefont {K.}~\bibnamefont
  {Hukushima}},\ }\href@noop {} {\bibfield  {journal} {\bibinfo  {journal}
  {Journal of Statistical Mechanics: Theory and Experiment}\ }\textbf {\bibinfo
  {volume} {2019}},\ \bibinfo {pages} {043102} (\bibinfo {year}
  {2019})}\BibitemShut {NoStop}%
\bibitem [{\citenamefont {Binder}(1981)}]{BC1}%
  \BibitemOpen
  \bibfield  {author} {\bibinfo {author} {\bibfnamefont {K.}~\bibnamefont
  {Binder}},\ }\href@noop {} {\bibfield  {journal} {\bibinfo  {journal} {Z.
  Phys. B Condens. Matter}\ }\textbf {\bibinfo {volume} {43}},\ \bibinfo
  {pages} {119} (\bibinfo {year} {1981})}\BibitemShut {NoStop}%
\bibitem [{\citenamefont {Jin}\ \emph {et~al.}(2012)\citenamefont {Jin},
  \citenamefont {Sen},\ and\ \citenamefont {Sandvik}}]{BCE}%
  \BibitemOpen
  \bibfield  {author} {\bibinfo {author} {\bibfnamefont {S.}~\bibnamefont
  {Jin}}, \bibinfo {author} {\bibfnamefont {A.}~\bibnamefont {Sen}}, \ and\
  \bibinfo {author} {\bibfnamefont {A.~W.}\ \bibnamefont {Sandvik}},\
  }\href@noop {} {\bibfield  {journal} {\bibinfo  {journal} {Phys. Rev. Lett.}\
  }\textbf {\bibinfo {volume} {108}},\ \bibinfo {pages} {045702} (\bibinfo
  {year} {2012})}\BibitemShut {NoStop}%
\bibitem [{\citenamefont {Binder}(1987)}]{BC2}%
  \BibitemOpen
  \bibfield  {author} {\bibinfo {author} {\bibfnamefont {K.}~\bibnamefont
  {Binder}},\ }\href@noop {} {\bibfield  {journal} {\bibinfo  {journal} {Rep.
  Progr. Phys.}\ }\textbf {\bibinfo {volume} {50}},\ \bibinfo {pages} {783}
  (\bibinfo {year} {1987})}\BibitemShut {NoStop}%
\bibitem [{\citenamefont {Weinberg}\ and\ \citenamefont {Bukov}(2017)}]{Phil}%
  \BibitemOpen
  \bibfield  {author} {\bibinfo {author} {\bibfnamefont {P.}~\bibnamefont
  {Weinberg}}\ and\ \bibinfo {author} {\bibfnamefont {M.}~\bibnamefont
  {Bukov}},\ }\href@noop {} {\bibfield  {journal} {\bibinfo  {journal} {SciPost
  Physics}\ }\textbf {\bibinfo {volume} {2}},\ \bibinfo {pages} {003} (\bibinfo
  {year} {2017})}\BibitemShut {NoStop}%
\end{thebibliography}%

\end{document}